\documentclass[journal]{IEEEtran}
\pdfoutput=1
\usepackage[dvipsnames,svgnames]{xcolor}
\usepackage{mathrsfs}
\usepackage{multirow}
\usepackage{makecell}
\usepackage{booktabs}
\usepackage{cite}
\usepackage{graphicx}
\usepackage{float}
\usepackage{url}
\usepackage{amsmath}
\usepackage{bm}

\usepackage{amsfonts}
\makeatletter
\newcommand{\removelatexerror}{\let\@latex@error\@gobble}
\makeatother
\usepackage{color}
\usepackage{algorithm}
\usepackage{algorithmic}
\usepackage{array}
\usepackage{booktabs}
\usepackage{multirow}
\usepackage{ragged2e}
\usepackage{amssymb}

\usepackage[T3,T1]{fontenc}
\DeclareSymbolFont{tipa}{T3}{cmr}{m}{n}
\DeclareMathAccent{\invbreve}{\mathalpha}{tipa}{16}

\ifCLASSOPTIONcompsoc
\usepackage[caption=false,font=normalsize,labelfont=sf,textfont=sf]{subfig}
\else
\usepackage[caption=false,font=footnotesize]{subfig}
\fi
\usepackage{stfloats}

\ifCLASSINFOpdf
\else
\fi

\begin{document}
	\title{Decomposing and Coupling Saliency Map for Lesion Segmentation in Ultrasound Images}
	\author{Zhenyuan~Ning, Yixiao~Mao, Qianjin~Feng, Shengzhou~Zhong, and Yu~Zhang
		\thanks{This work has been submitted to the IEEE for possible publication. Copyright may be transferred without notice, after which this version may no longer be accessible. This work was supported in part by the National Natural Science Foundation of China under Grant U22A20350 and Grant 61971213, in part by the Basic and Applied Basic Research Foundation of Guangdong Province under Grant 2019A1515010417, and in part by the Guangdong Provincial Key Laboratory of Medical Image Processing under Grant No.2020B1212060039. The approvals of the Ethics
		Committee are not available, in that the datasets used are obtained from public databases, and we have cited the required references. (Zhenyuan Ning and Yixiao Mao contributed equally to this work.) (Corresponding author: S. Zhong and Y. Zhang).
			
		The authors are with the School of Biomedical Engineering, Southern Medical University, Guangzhou, Guangdong, 510515, China (E-mail: jonnyning@foxmail.com; mao0823xiao@foxmail.com; fengqj99@smu.edu.cn; shengzhouzhong@foxmail.com; yuzhang@smu.edu.cn). 
		They are also with the Guangdong Provincial Key Laboratory of Medical Image Processing and the Guangdong Province Engineering Laboratory for Medical Imaging and Diagnostic Technology, Southern Medical University, Guangzhou 510515, China.
	}}
	\maketitle
	
	\begin{abstract}
		
		Complex scenario of ultrasound image, in which adjacent tissues (i.e., background) share similar intensity with and even contain richer texture patterns than lesion region (i.e., foreground), brings a unique challenge for accurate lesion segmentation.
		This work presents a decomposition-coupling network, called DC-Net, to deal with this challenge in a (foreground-background) saliency map disentanglement-fusion manner.
		The DC-Net consists of decomposition and coupling subnets, and the former preliminarily disentangles original image into foreground and background saliency maps, followed by the latter for accurate segmentation under the assistance of saliency prior fusion.
		The coupling subnet involves three aspects of fusion strategies, including: 
		1) regional feature aggregation (via differentiable context pooling operator in the encoder) to adaptively preserve local contextual details with the larger receptive field during dimension reduction; 
		2) relation-aware representation fusion (via cross-correlation fusion module in the decoder) to efficiently fuse low-level visual characteristics and high-level semantic features during resolution restoration;
		3) dependency-aware prior incorporation (via coupler) to reinforce foreground-salient representation with the complementary information derived from background representation.
		Furthermore, a harmonic loss function is introduced to encourage the network to focus more attention on low-confidence and hard samples.
		The proposed method is evaluated on two ultrasound lesion segmentation tasks, which demonstrates the remarkable performance improvement over existing state-of-the-art methods.
		
	\end{abstract}
	\begin{IEEEkeywords}
		Lesion segmentation, ultrasound image, saliency map, complex scenario, deep learning
	\end{IEEEkeywords}
	
	\section{Introduction}
	\IEEEPARstart{U}ltrasound imaging, as one of the most common imaging schemes in clinical practices, has been extensively applied to the early detection of many diseases (e.g., breast cancer and thyroid nodule) in view of its safety and efficiency~\cite{noble2006ultrasound, huang2018machine}.
	Clinically, radiologists can make preliminary diagnosis by screening ultrasound images, which is experience-dependent and may suffer from high inter-observer variation even for well-trained radiologists~\cite{james2014medical}.
	For decades, computer-aided diagnosis (CAD) system based on ultrasound images is developed to help radiologists improve diagnostic accuracy and reduce individual subjectivity~\cite{daoud2019automatic}, in which lesion segmentation acts as a significant step towards sensitivity and efficiency improvement~\cite{sloun2019deep}.
	Therefore, it has attracted much attention on automatic and robust segmentation methods for advancing the development of CAD system~\cite{huang2018machine,qureshi2022medical}.
	
\begin{figure}[t]
		\includegraphics[width=1\linewidth]{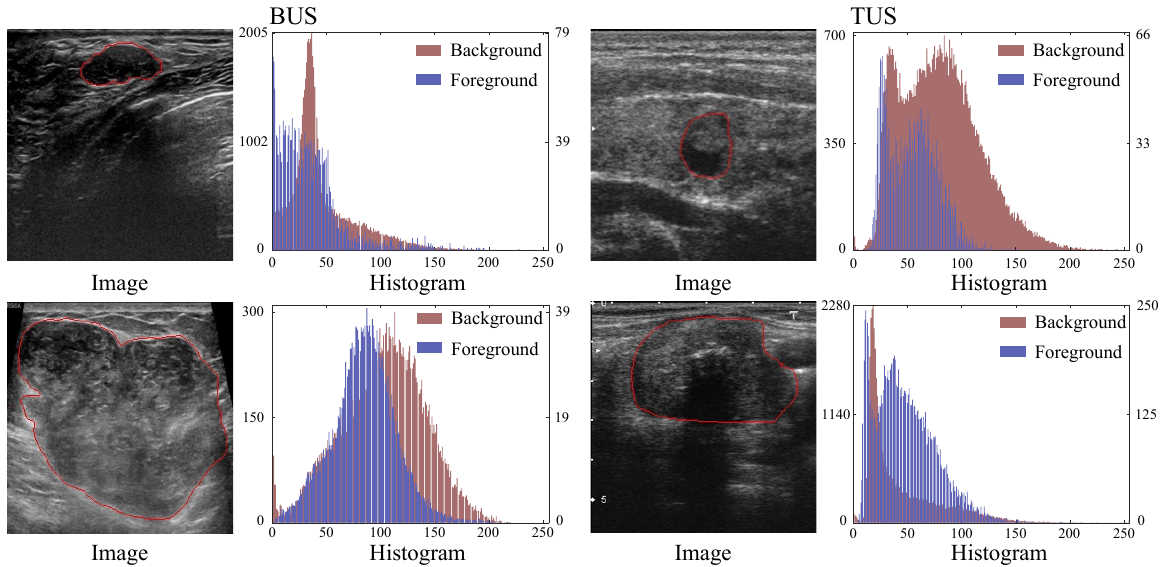}
		\caption{The histogram statistics of representative cases from two ultrasound image databases. The x-axis denotes the intensity, while y-axis represents the number count (left: background, right: foreground).}
		\label{fig_1}
		\vspace{-5mm}
	\end{figure}

	In recent years, many segmentation methods have been proposed for automatic lesion delineation in ultrasound images, among which deep learning methods, especially convolutional neural networks (CNNs), have showed the superior performance to most of conventional models (e.g., model-based and region-based approaches)~\cite{xian2018automatic, clough2020topological}.
	However, complex scenario still poses a unique challenge for accurate lesion segmentation~\cite{shao2015saliency, liu2019deep}.
	For intuitive illustration, Fig.~\ref{fig_1} exhibits the histogram statistics of several representative cases from two ultrasound image databases.
	We can observe that lesion regions (i.e., foreground) generally show low-intensity signals, while the surrounding tissues (i.e., background) share similar intensity profiles with and even contain richer texture patterns than foreground, making it difficult to accurately separate foreground from background.
	In consideration of such scenario, recent studies have tried the utilization of the cues provided by saliency maps to boost model's generalization performance on foreground segmentation~\cite{wang2018deepigeos, vakanski2020attention, ning2021smu, chu2021ultrasonic}.
	These methods commonly cluster and highlight visually salient regions to assist subsequent segmentation models, which, however, is a standalone preprocessing step that requires the prior of interactive hand-clicked seeds.
	A desirable but rare way is to deal with this challenge in a (foreground-background) saliency map disentanglement-fusion manner, in which saliency map generation and lesion segmentation are conducted in a collaborative manner without manual intervention.
	
	In general, the vast majority of segmentation networks on ultrasound images, as the variants of fully convolutional network (FCN)~\cite{long2015fully} and U-Net~\cite{ronneberger2015unet}, equip with a contracting path (i.e., encoder) for morphological context extraction as well as a symmetric expanding path (i.e., decoder) for precise localization and segmentation~\cite{long2015fully, ronneberger2015unet}.
	However, the performance gain of those methods is hindered by insufficient information fusion.
	For encoder, it typically utilizes the pooling layer to aggregate regional features so as to reduce feature dimension and enlarge receptive field. 
	Although the pooling layer can efficiently alleviate the overfitting issue and help capture global information, it inevitably results in information dropout especially on relatively small lesions.
	As to decoder, the skip connectivity operator is exploited to transmit morphological features from encoding path to decoding one for preserving the low-level characteristics of images, which commonly ignores the inner association during the fusion process of low-level visual information and high-level semantic representation.
	On the other hand, most of existing methods may prefer to optimize the segmentation of easy samples with distinct edge and appearance during the training stage, but inattention to low-confidence and hard samples limits the generalisation of models.
	We argue that the network can work better on ultrasound lesion segmentation task when it can deal with the aforementioned issues.
	
	In this paper, we propose a decomposition-coupling network (DC-Net) to incorporate saliency map generation into the model for ultrasound lesion segmentation.
	Decomposition and coupling subnets jointly constitute to form DC-Net, in which the former preliminarily disentangles the original image into foreground and background saliency maps, while the latter conducts accurate lesion segmentation by coupling the saliency prior of foreground with that of background.
	The coupling subnet consists of an encoder, a decoder and a coupler, involving three aspects of fusion strategies: 
	1) The encoder is armed with a differentiable context pooling strategy to aggregate regional features, which can adaptively preserve local contextual details with the larger receptive field during dimension reduction. 
	2) The decoder conducts relation-aware representation fusion of low-level visual characteristics and high-level semantic features during resolution restoration. 
	3) A coupler is embedded to reinforce foreground-salient representation by incorporating the complementary information derived from background representation based on its dependency on foreground.
	Furthermore, we present a harmonic loss function for extra performance improvement by compelling the network to focus on low-confidence samples in addition to hard ones.
	The main contributions of this paper lie in the following aspects:
	\begin{itemize} 	
		\item[--] We try a saliency map disentanglement-fusion manner for lesion segmentation in complex-scenario ultrasound images by the introduction of DC-Net which simultaneously conducts saliency map generation and lesion segmentation without manual intervention.
		\item[--] We devise a coupling subnet, involving three aspects of fusion strategies, including regional feature aggregation (via differentiable context pooling operator in the encoer), relation-aware representation fusion (cross-correlation fusion module  in the decoder) and dependency-aware prior incorporation (via coupler). 
		\item[--] We propose a harmonic loss function to prevent low-confidence samples as well as hard ones from being overwhelmed during the training stage by lifting or keeping their gradients. Interestingly, the harmonic loss can degenerate into the focal loss and cross-entropy loss under certain conditions.
		\item[--] Extensive experiments on two ultrasound lesion segmentation tasks (i.e., breast ultrasound segmentation and thyroid ultrasound segmentation tasks) demonstrate its superiority.
	\end{itemize}
	
	The remaining portion of this paper is organized as follows.
	In Section~\ref{RelatedWork}, we briefly review the related work.
	We detail our specific method in Section 3.
	Section 4 presents experimental settings and results. The discussion and conclusion are drawn in Section 5 and 6, respectively.
	
	\section{Related Work}
	\label{RelatedWork}
	\subsection{Saliency Prior-based Segmentation Approaches} 
	Many studies have recently demonstrated the potential of the cues provided by saliency prior to help improve model’s generalization performance~\cite{minaee2021image, cheng2021modular, nieuwenhuis2012spatially}, especially on medical image segmentation tasks~\cite{feng2021interactive, cho2021deepscribble}. These methods commonly cluster and highlight visually salient regions to assist downstream segmentation models.
	For example, Luo et al.~\cite{luo2021mideepseg} considered the exponentialized geodesic distance as saliency map and took it as additional input of the network.
	Feng et al.~\cite{feng2021interactive} utilized the saliency cue from interactive manual correction on the prediction map in an iterative manner.
	Vakanski et al.~\cite{vakanski2020attention} generated the visual saliency map and integrated it into the model via attention mechanism for accurate lesion segmentation.
	However, most methods only focus on foreground saliency prior and ignore the structural characteristics contained by background, which may result in sub-optimal solutions especially in the complex scenario of ultrasound images.
	Actually, several studies have manifested that the introduction of background prior also contributes to the performance gain~\cite{xu2016deep, wang2018deepigeos, zhang2021dins, ning2021smu}.
	For instance, Xu et al.~\cite{xu2016deep} transformed user-provided positive and negative clicks into foreground and background saliency maps, for boosting the network's capability of learning target-related representation.
	Wang et al.~\cite{wang2018deepigeos} proposed to form the geodesic distance maps of foreground and background based on the initial prediction map and concatenate them with the original image as model's input for refining segmentation results.
	Ning et al.~\cite{ning2021smu} tried to generate hierarchical foreground and background saliency maps that were hinted by three random seeds to assist downstream lesion segmentation model.
	However, existing methods generally neglect the potential associations between foreground and background, which may lead to inadequate exploitation of complementary information.
	Additionally, most methods adopt standalone generation algorithms that usually require interactive hand-clicked seeds to produce saliency prior, which, however, is time-consuming and experience-dependent.
	In other word, the learning of saliency map and segmentation model is carried out in two separated processes and it may lead to a sub-optimal result, even though each of these two processes could achieve their individual optimization.
	An appealing way is to integrate saliency map generation and lesion segmentation into a unified framework without manual intervention, and exploit complementary information derived from background representation based on its dependency on foreground to reinforce foreground-salient representation learning for accurate lesion segmentation.
	
	\subsection{Ultrasound Lesion Segmentation Networks}
	Convolutional neural networks have revolutionized ultrasound lesion segmentation over the last few years~\cite{avola2021ultrasound, komatsu2021towards,wang2021advances}, most of which belong to the variants of FCN and U-Net.
	For instance, Wang~\cite{wang2020automatic} et al. stacked two DeeplabV3plus to perform coarse-to-fine thyroid nodule segmentation and obtained promising results.
	Yap et al.~\cite{yap2017automated} developed several FCN-based variants for the semantic segmentation of breast lesion.
	However, existing methods are still confronted with some difficulties in feature fusion during dimension reduction and resolution restoration.
	For the former, pooling operator in the encoder is commonly used to aggregate neighborhood features for reducing spatial resolution and enlarging receptive field.
	For lesion segmentation networks, default pooling operators include average pooling~\cite{pan2021sgunet}, max pooling~\cite{gong2021multi, lee2020channel} and stride convolution due to their widespread adoption and proven efficacy~\cite{zou2021robust}.
	Although these pooling operators are straightforward to aggregate regional features and alleviate the overfitting issue, they may also result in detail information dropout especially on relatively small lesions. 
	Currently, some works have suggested that local detail preservation during the pooling stage is beneficial to model's generalization~\cite{gao2019lip, saeedan2018detail}.
	For example, Saeedan et al.~\cite{saeedan2018detail} proposed a detail-preserving pooling strategy to magnify spatial changes and reserve structural patterns.
	Gao et al.~\cite{gao2019lip} devised a local importance-based pooling operator to aggregate features within the window for detail preservation.
	However, these methods only concentrate on feature aggregation at the specified pooling window with a fixed receptive field, neglecting the significant contextual details inherent in neighborhoods (naturally with larger receptive field).
	As to the feature fusion during resolution restoration, skip connectivity mechanism~\cite{drozdzal2016importance} (commonly with channel-wise concatenation \cite{ning2021smu, RDAU-NET, vakanski2020attention, byra2020breast, CF2-Net, shahroudnejad2021thyroid, wu2020ultrasound, pan2021sgunet} or pixel-wise addition~\cite{yang2021efficient}) is introduced to transmit the morphological context features from encoding blocks to decoding ones.
	To promote the efficiency of feature transmission and fusion, some works have ameliorated the skip connection strategy~\cite{liu2020end,honghan2021rms}.
	For example, Liu et al.~\cite{liu2020end} modified U-Net with the redesigned skip connection operator to merge multi-scale semantic features for thyroid nodule stratification.
	Zhu et al.~\cite{honghan2021rms} introduced a multi-skip-connection strategy that cooperated with the squeeze-and-excitation block to fuse the feature maps from encoder and decoder.
	Nevertheless, most of them ignore the inner relationship between low-level visual information and high-level semantic representation, and previous studies have demonstrated that feature fusion based on correlation would reduce feature redundancy and is conducive to feature propagation~\cite{Unet++, han2020ghostnet, wang2021tied}.
	
	As another challenge for ultrasound image segmentation, the hard or low-confidence samples/pixels (around blur edge and appearance) generally occupy a smaller proportion than the easy ones, which means the easy samples would dominate gradient descent during the training stage and might result in sub-optimal performance~\cite{zheng2020foreground}.
	To this end, some researches have designed supervision functions based on focal loss for emphasizing on training a sparse set of hard samples~\cite{lin2020focal,li2021multi,yeung2022unified}.
	For example, Li et al.~\cite{li2021multi} proposed to combine focal loss and dice loss to promote the model's capability of segmenting hard samples.
	Yeung et al.~\cite{yeung2022unified} presented a unified focal loss that generalized dice loss and cross-entropy loss for coping with hard samples.
	Compared with cross-entropy (CE) loss, focal loss introduces a factor $(1-p)^\gamma$ to adjust the loss assigned to samples, which relatively raises the network's concern for hard samples. 
	However, as $\gamma$ increases, focal loss further down-weights easy samples, but concurrently overwhelms some relatively low-confidence ones (e.g., $p=.4\sim.75$). 
	Actually, those samples still have positive effects on network training, which might be one of the reasons for the best choice of small $\gamma$ in most methods~\cite{lin2020focal, li2021multi}.
	Therefore, we advocate to enforce the network to simultaneously focus on hard and low-confidence samples.
	
	\section{Methods}
	\subsection{Problem Formulation}
	We aim to distinguish lesion region (foreground) from its surrounding tissues (background) in ultrasound images.
	Denote $\mathcal{D}=\{\textbf{I}_i, \textbf{Y}_i\}_{i=1}^n$ as a dataset containing $n$ instances, where \textbf{I}$_i$ and \textbf{Y}$_i$ represent the $i$-th original image and its corresponding pixel-wise annotation, respectively. An automated segmentation model $\mathbb{C}$ can be formulated as
	\begin{equation}
		\label{eq_seg}
		\hat{\textbf{Y}}_i = \mathbb{C}{(\textbf{I}_i)},
	\end{equation}
	where $\hat{\textbf{Y}}_i$ denotes the estimated label for the input image.
	
\begin{figure*}[t]
		\includegraphics[width=1\linewidth]{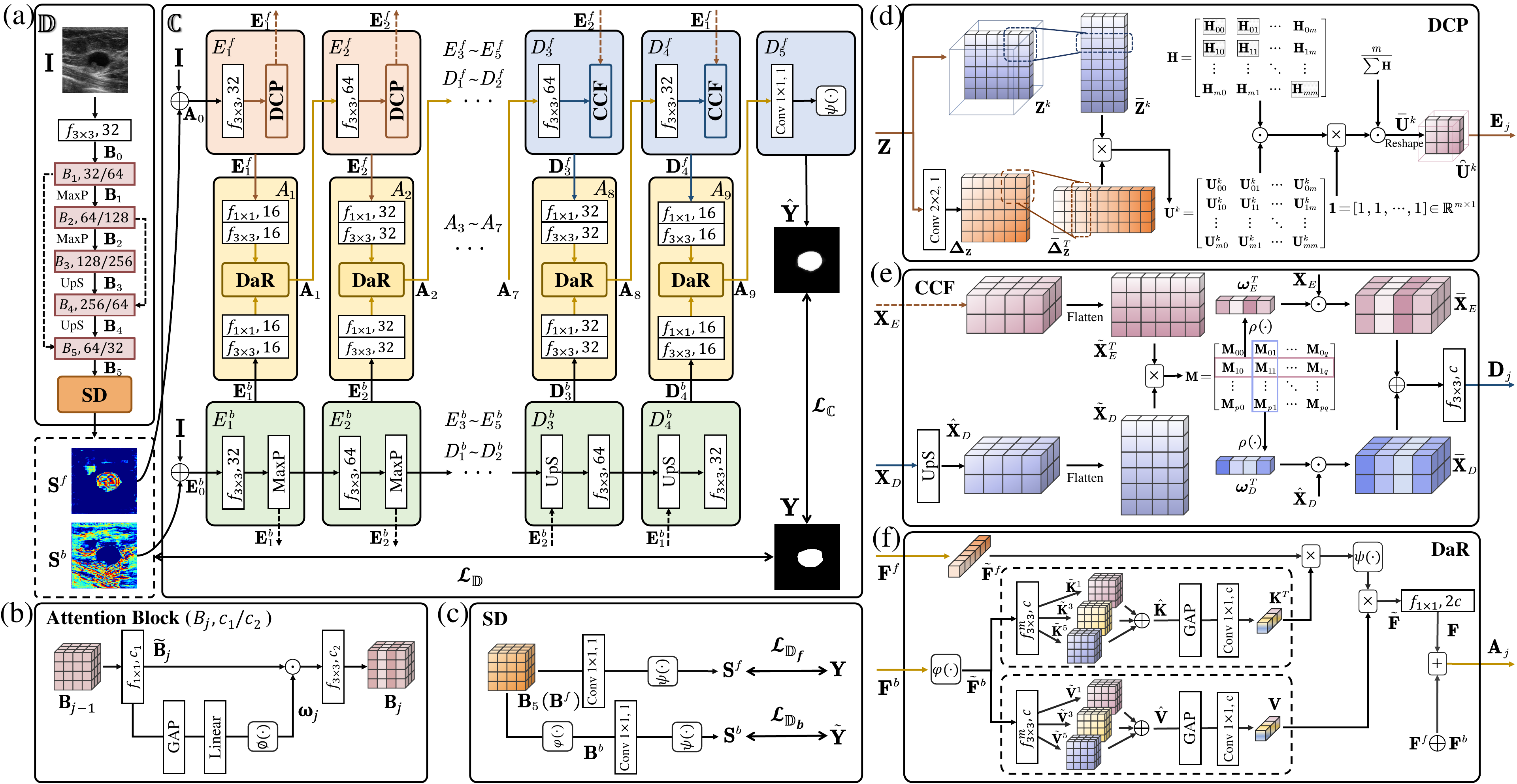}
		\caption{Overview of DC-Net. (a) Architecture of DC-Net. It consists of decomposition and coupling subnets. The decomposition subnet includes an attention backbone $\color{Thistle} \blacksquare$ (implemented by a convolutional unit $f_{3 \times 3}$ and five attention blocks $\{B_j\}_{j=1}^5$) and a saliency decomposition (SD) module $\color{BurntOrange} \blacksquare$. The coupling subnet involves an encoder $\color{Apricot} \blacksquare$ (constructed by five encoding blocks $\{E_j^f\}^5_{j=1}$), a decoder $\color{CornflowerBlue} \blacksquare$ (built by five decoding blocks $\{D_j^f\}^5_{j=1}$) and a coupler (constituted by an auxiliary stream $\color{SpringGreen} \blacksquare$, i.e., five contracting blocks $\{E_j^b\}^5_{j=1}$ and four expanding blocks $\{D_j^b\}^4_{j=1}$, and an aggregation stream $\color{Goldenrod} \blacksquare$, i.e., nine aggregation units $\{A_j\}_{j=1}^9$). (b) Illustration of attention block $B_j$. (c) Illustration of SD module. (d) Illustration of differentiable context pooling (DCP) module.  (e) Illustration of cross-correlation fusion (CCF) module. (f) Illustration of dependency-aware reinforcement (DaR) module. Notably, "MaxP", "UpS", "GAP" and "Linear" denote max pooling, upsampling, global average pooling and linear layer, respectively. $\phi(\cdot)$, $\varphi(\cdot)$ and $\psi(\cdot)$ represent softmax activation, self-reversal activation and sigmoid activation, respectively. Dotted arrow denotes skip connection operation.}
		\label{fig_2}
		\vspace{-5mm}
	\end{figure*}

	In practice, it is difficult to construct a satisfactory segmentation model only based on original ultrasound images due mainly to blur boundary and appearance encountered in the complex scenario.
	Thus, recent studies have sought help from foreground and background saliency maps~\cite{wang2018deepigeos, ning2021smu, zhang2021dins}.
	Accordingly, given $\mathcal{S}^f=\{\textbf{S}_i^f\}_{i=1}^n$ and $\mathcal{S}^b=\{\textbf{S}_i^b\}_{i=1}^n$ as foreground and background saliency maps, respectively, the segmentation model $\mathbb{C}$ is learned based on the saliency maps in addition to the original images, which is formulated as
	\begin{equation}
		\label{eq_seg2}
		\hat{\mathbb{C}} = \mathop{\arg\min}\limits_{\mathbb{C}} \sum_{i=1}^n[\mathbb{C}{(\textbf{I}_i, \textbf{S}^f_i, \textbf{S}^b_i), \textbf{Y}_i}]_{\mathcal{L}},
	\end{equation}
	where $\mathcal{L}$ denotes the loss function.
	Generally, the generation of saliency maps requires the interactive hand-clicked seeds, which is time-consuming and experience-dependent.
	To address this issue, we first try to conduct lesion segmentation in a saliency map disentanglement-fusion manner and integrate saliency map generation and lesion segmentation into a collaborative framework without manual intervention.
	Let $\mathbb{D}_f$ and $\mathbb{D}_b$ represent the mapping functions that transform the original image into saliency maps, i.e., $\textbf{S}^f_i=\mathbb{D}_f(\textbf{I}_i)$ and $\textbf{S}^b_i=\mathbb{D}_b(\textbf{I}_i)$.
	Then, Eq.(\ref{eq_seg2}) can be rewritten as
	\begin{equation}
		\label{eq_seg3}
		{\hat{\mathbb{C}},\hat{\mathbb{D}}_*}=\mathop{\arg\min}\limits_{\mathbb{C}, \mathbb{D_*}} \sum_{i=1}^n[\mathbb{C}{(\textbf{I}_i, \mathbb{D}_f(\textbf{I}_i), \mathbb{D}_b(\textbf{I}_i)), \textbf{Y}_i}]_{\mathcal{L}},
	\end{equation}
	where $*$ denotes $f$ and $b$.
	Based on Eq.(\ref{eq_seg3}), we propose DC-Net which includes two  major components, namely a decomposition subnet (i.e., $\mathbb{D}$) for saliency map generation and a coupling subnet (i.e., $\mathbb{C}$) for lesion segmentation with the assistance of saliency prior fusion, as illustrated in Fig.~\ref{fig_2}(a).
	Notably, the decomposition subnet $\mathbb{D}$ is utilized to produce foreground and background saliency maps simultaneously.
	
	\subsection{Decomposition Subnet}
	Different from previous methods that utilize manual intervention to generate saliency prior~\cite{wang2018deepigeos, ning2021smu}, the decomposition subnet $\mathbb{D}$ (Fig.~\ref{fig_2}(a)) disentangles the original image into foreground and background saliency maps in a data-driven fashion. It is primarily composed of an attention backbone which contains a convolutional unit and five attention blocks (Fig.~\ref{fig_2}(b)), and a saliency decomposition module (Fig.~\ref{fig_2}(c)).
	
	\subsubsection{Attention Backbone}
	Given an input image $\textbf{I}$, a convolutional unit $f_{3 \times 3}$\footnote{Unless otherwise specified, the convolutional unit used in this paper, which consists of an $a \times a$ convolutional layer, a batch normalization layer and a ReLU activation operation, is denoted as $f_{a \times a}$.} is first utilized to mine shallow feature $\textbf B_0$ from $\textbf{I}$.
	Then, five attention blocks $\{B_j\}_{j=1}^5$ with the same architecture are inserted for learning region-salient representations.
	For the $j$-th block $B_j$, it takes \textbf{B}$_{j-1}$ (i.e., $\textbf B_0$ or the output of $B_{j-1}$) as input and generates the feature $\tilde{\textbf{B}}_j$ via a $f_{1 \times 1}$.
	Subsequently, a global average pooling layer and a fully-connected layer are carried on $\tilde{\textbf{B}}_{j}$ to obtain channel-wise weight vector $\bm \omega _j$ that is activated by a softmax function $\phi$.
	After that, $\bm \omega _j$ is adopted to enhance the significant channels of $\tilde{\textbf{B}}$$_{j}$, which is further merged by another $f_{3 \times 3}$ to get the representation $\textbf{B}_j$:
	\begin{equation}
		\label{eq_att2}
		\textbf{B}_j=f_{3 \times 3}(\bm \omega _j \odot \tilde{\textbf{B}}_j),
	\end{equation}
	where $\odot$ denotes the channel-wise multiplication operation. 
	Note that all blocks $\{B_j\}_{j=1}^5$ contain two convolutional layers and their channel numbers are empirically set to $\{32/64\}$, $\{64/128\}$, $\{128/256\}$, $\{256/64\}$ and $\{64/32\}$, respectively.
	And two $4\times 4$ max-pooling layers are respectively placed behind the first two blocks for resolution reduction, while two $4\times 4$ up-sampling layers in front of the last two blocks for resolution restoration, as shown in Fig.~\ref{fig_2}(a-b).
	Meanwhile, the skip connectivity operator is used to concatenate the output features of $B_1$/$B_2$ and the input features of $B_5$/$B_4$.
	
	\subsubsection{Saliency Decomposition Module}
	A saliency decomposition (SD) module is constructed for the generation of foreground and background saliency maps.
	In this module, as shown in Fig.~\ref{fig_2}(c), we first apply a self-reversal activation operator $\varphi$ to \textbf{B}$_5$ (also named as \textbf{B}$^f$ for convenience), so as to obtain complementary background representation \textbf{B}$^b$:
	\begin{equation}
		\label{eq_att3}
		\textbf{B}^b=\varphi(\textbf{B}^f)=\{max\{{\textbf B}^{f, k}\}-{\textbf B}^{f, k}\}^c_{k=1},
	\end{equation}
	where $c$ denotes the channel number of \textbf{B}$^f$. 
	Then, a $1 \times 1$ convolutional layer and a sigmoid activation function $\psi$ are plugged to get foreground and background saliency maps (i.e., $\textbf{S}^f$ and $\textbf{S}^b$).
	Furthermore, we adopt a weakly-supervised manner, that is using the foreground mask $\textbf{Y}$ and background mask $\tilde{\textbf{Y}} = \varphi(\textbf{Y})$ as collaborative supervision signals rather than complex manual saliency prior, to train the network.
	Consequently, based on the dataset $\mathcal{D}=\{\textbf{I}_i, \textbf{Y}_i\}_{i=1}^n$, it can be formulated as
	\begin{equation}
		\label{eq_att4}
		\mathop{\min}\limits_{\mathbb{D}} \sum_{i=1}^n\{[\textbf{S}^f_i, \textbf{Y}_i]_{\mathcal{L}_{\mathbb{D}_f}}+[\textbf{S}^b_i, \tilde{\textbf{Y}}_i]_{\mathcal{L}_{\mathbb{D}_b}}\},
	\end{equation}
	where $\mathcal{L}_{\mathbb{D}_f}$ and $\mathcal{L}_{\mathbb{D}_b}$ denote the loss functions for learning foreground and background saliency maps, respectively.
	\\\noindent{\textbf{Remark 1:}}
	\textit{This restricted form of previous saliency map generation methods limits their generality and usability since additional interactive knowledge is needed to specify visual focal points.  By comparison, disentangling directly from original image about saliency maps is a promising alternative which is first work we are aware of that has been integrated into the downstream segmentation model and optimized jointly. }
	
	\subsection{Coupling Subnet}
	The coupling subnet $\mathbb{C}$ is proposed for accurate lesion segmentation under the assistance of saliency maps, consisting of an encoder, a decoder and a coupler.
	Unlike prior ultrasound lesion segmentation networks \cite{ning2021smu,RDAU-NET,byra2020breast,shahroudnejad2021thyroid,wu2020ultrasound,pan2021sgunet}, the coupling subnet involves three fusion modifications: 
	1) A differentiable context pooling (DCP) operator is devised and inserted in the encoder to aggregate regional features and adaptively preserve local contextual details during dimension reduction;
	2) The decoder is armed with a cross-correlation fusion (CCF) module to model the relation between low-level and high-level features and fuse them during resolution restoration;
	3) A coupler is embedded to reinforce foreground-salient representation learning by incorporating complementary information derived from background-salient cues based on its dependency on foreground.
	
	\subsubsection{{Encoder}}
	In Fig.~\ref{fig_2}(a), the encoder is constructed based on five blocks $\{E_j^f\}^5_{j=1}$, and all blocks share the same architecture except for the block $E_5^f$ that removes the pooling operator.
	For the $j$-th block $E_j^f$, it intakes ${\textbf A}_{j-1}$ (i.e., ${\textbf A}_{0}$\footnote{We denote the concatenation of original image and foreground saliency map as ${\textbf A}_{0}$, namely ${\textbf A}_{0}={\textbf I} \oplus \textbf{S}^f$.} or the output of the aggregation unit ${A}_{j-1}$).
	Then, a convolutional unit $f_{3 \times 3}$ is used to extract features, followed by a DCP operator to aggregate regional features, shrink the spatial size of the feature maps and obtain the representation ${\textbf{E}}_j^f$. 
	
	\textit{\textbf{$\rhd$ DCP Operator}}:
	In general, most pooling methods can be regarded as a convex combination of local neuron activations~\cite{kobayashi2019gaussian}. 
	Given the $k$-th feature map ${\textbf{Z}}^k$ from $\textbf{Z} \in \mathbb{R} ^{h\times {w}\times {c}}$, such pooling operation can be formulated as
	\begin{small}
		\begin{equation}
			\label{eq_dcp1}
			\invbreve{\textbf{Z}}^{k}=\left\{ \sum_{\boxplus_p}{\textbf{Z}^{k}_{\boxplus_p} \odot \bm{\delta}_{\boxplus_p}} \right\} _{p=1}^{m},\\
		\end{equation}
	\end{small}%
	where $\textbf{Z}^{k}_{\boxplus_p} \in \mathbb{R}^{r \times r}$ and $\bm{\delta}_{\boxplus_p}\in \mathbb{R}^{r \times r}$ represent the features and pooling kernel within the $p$-th sliding window $\boxplus_p$, respectively, and $m$ denotes the number of sliding windows and is computed by $\frac{h}{r} \times \frac{w}{r}$.
	Generally speaking, ${\bm{\delta}_{\boxplus_p}}$ is non-learnable and sharable for all windows.
	However, Eq.(\ref{eq_dcp1}) only concentrates on the specified window with a fixed receptive field, neglecting the contextual information inherent in neighborhoods.
	Recent work has demonstrated that contexts enable the network to retain essential details and understand image structure in ultrasound images \cite{ning2021smu}. 
	To this end, we rethink the pooling operator from the perspective of matrix manipulation. 
	We first define the pooling kernel as a matrix $\mathbf \Delta \in \mathbb{R} ^{h\times {w}}$ that contains $m$ non-sharable pooling kernels, i.e., $\{{\bm{\delta}_{\boxplus_p}}\}_{p=1}^{m}$. 
	Then, as shown in Fig.~\ref{fig_2}(d), ${\textbf{Z}}^k$ and $\mathbf \Delta$ are flattened based on the window and stacked as $\bar{\textbf{Z}}^k\in \mathbb{R} ^{m\times r^2}$ and $\mathbf{\bar{\Delta}}\in \mathbb{R} ^{m\times r^2}$, respectively. 
	We further project $\bar{\textbf{Z}}^k$ into a latent space $\textbf{U}^k \in \mathbb{R}^{m \times m}$ by $\mathbf{\bar{\Delta}}$:
	\begin{equation}
		\label{eq_dcp2}
		\textbf{U}^k=\bar{\textbf{Z}}^k\times \mathbf{\bar{\Delta}}^T,
	\end{equation}
	where the entry $\textbf{U}^k_{p,q}$ ($p,q\in\{1,...,m\}$) denotes the representation that is obtained by using the $q$-th kernel to map the features within the $p$-th window.
	Considering that contexts generally appear in local neighborhoods, we introduce an indicator matrix $\mathbf{H} \in \mathbb{R}^{m \times m}$ to activate $\mathbf{U}^k$ such as to extract and aggregate contextual details (to generate $\bar{\textbf{U}}^k \in \mathbb{R}^{m \times 1}$) as follows:\\
	\begin{small}
		\begin{equation}
			\label{eq_dcp3}
			\begin{split}
				\bar{\textbf{U}}^k&=\frac{m}{\sum{\textbf{H}}} (\textbf{H} \odot \textbf{U}^k) \times \textbf{1}
				=\frac{m}{\sum{\textbf{H}}} (\textbf{H} \odot (\bar{\textbf{Z}}^k\times \mathbf{\bar{\Delta}}^T))\times \textbf{1},
			\end{split}
		\end{equation}
	\end{small}%
	where $\textbf{1} \in \mathbb{R}^{m \times 1}$ is an all-ones vector, $\frac{m}{\sum{\textbf{H}}}$ is a smooth factor, and $\textbf{H}$ is essentially a band matrix and has the following definition.
	\\$\textbf{Definition 1 (Indicator Matrix $\mathbf{H}$).}$ 
	\textit{Given the ${m \times m}$ matrix $\mathbf{H}=(\mathbf{H}_{p,q})$ $(p,q\in\{1,...,m\})$, all entries are 0 outside a diagonally bordered band whose range is determined by constant $\tau (\geq 0)$, and the rest is equal to 1. More formally, it can be defined as}
	\begin{equation}
		\mathbf{H}_{p,q} = 
		\begin{cases}
			0, &  \text {if} \quad q < p - \tau \vee q > p + \tau \\  1, & \text{otherwise} \end{cases}.
	\end{equation}
	Through $\textbf{H}$,  Eq.(\ref{eq_dcp3}) can on one hand preserve local contextual details and on the other hand enlarge receptive field when neighborhoods are referred.
	For instance, if $\textbf{H}$ is a tridiagonal matrix, it associates eight neighborhoods of a window and the receptive field is concomitantly enlarged to eight times as against conventional pooling strategies. 
	Intuitively, when $\textbf{H}$ is an identity matrix (i.e., neglecting contexts), we can obtain two special cases of Eq.(\ref{eq_dcp3}) as follows:
	i) if $\mathbf{\bar{\Delta}}$ is a constant matrix and its entry $\mathbf{\bar{\Delta}}_{\mu,\nu}$ (${\mu}\in\{1,...,m\}, {\nu}\in\{1,...,r^2\}$) equals to $\frac{1}{r^2}$, Eq.(\ref{eq_dcp3}) is  an average pooling operator in essence;
	ii) if $\mathbf{\bar{\Delta}}$ is binary and $\mathbf{\bar{\Delta}}_{\mu,\nu}$ is equal to 1 in the condition that $\bar{\textbf{Z}}^k_{\mu,\nu}$ has the largest value in the $\mu$-th row (i.e., the $\mu$-th window in ${\textbf{Z}^k}$), Eq.(\ref{eq_dcp3}) can be considered as a max pooling operator.
	Additionally, as a remedy, it is expected that $\mathbf{\bar{\Delta}}$ is learnable and adaptive for input features.
	Consequently, we introduce a differentiable pooling kernel $\mathbf{\bar{\Delta}}_{\textbf{Z}}$ that can be learned in a task-driven manner, and thus Eq.(\ref{eq_dcp3}) can be reformulated as
	\begin{equation}
		\label{eq_dcp5}
		\begin{split}
			\bar{\textbf{U}}^k=\frac{m}{\sum{\textbf{H}}} (\textbf{H} \odot (\bar{\textbf{Z}}^k\times \mathbf{\bar{\Delta}}^T_{\textbf{Z}}))\times \textbf{1},
		\end{split}
	\end{equation}
	where $\mathbf{\bar{\Delta}}_{\textbf{Z}}$ is implemented by applying a $2 \times 2$ convolutional layer to $\textbf{Z}$.
	Finally, we reshape $\bar{\textbf{U}}^k$ to form $\invbreve{\textbf{U}}^{k}$ (i.e., $\mathbb{R}^{m \times 1} \mapsto \mathbb{R}^{\frac{h}{r} \times \frac{w}{r}}$) that is used as the output of DCP operator.
	\\\noindent{\textbf{Remark 2:}}
	\textit{We formulate DCP operator from the perspective of matrix manipulation, which generalizes conventional pooling approaches to aggregate regional features.
		This operator enables the network to adaptively preserve local contextual details to alleviate information dropout issue while enlarging receptive field.}
	
	\subsubsection{{Decoder}}
	Fig.~\ref{fig_2}(a) shows the architecture of decoder that includes five blocks $\{D_j^f\}_{j=1}^5$.
	For the first four blocks, each of them consists of a convolutional unit $f_{3 \times 3}$ and a cross-correlation fusion (CCF) module for resolution restoration, while the block $D_5^f$ involves a $1 \times 1$ convolutional layer and a sigmoid function $\psi$ for segmentation map estimation.
	In the $j$-th block $D_j^f$ (where $j \in \{1,\cdots,4\}$), a convolutional unit $f_{3 \times 3}$ is applied to $\textbf{A}_{4+j}$ to generate $\tilde{\textbf{D}}_j^f$, and then a CCF module is developed to obtain the up-sampled representation $\textbf{D}_j^f$ by fusing the low-level visual characteristic $\textbf{E}_{5-j}^f$ and high-level semantic representation $\tilde{\textbf{D}}_j^f$.
	For the block $D_5^f$, it receives $\textbf{A}_{9}$ and outputs $\hat{\textbf{Y}}$ to compute the loss to the annotation ${\textbf{Y}}$ for the network training. 
	Given the dataset $\mathcal{D}=\{\textbf{I}_i, \textbf{Y}_i\}_{i=1}^n$, it can be formulated as
	\begin{equation}
		\label{eq_Lc}
		\begin{split}
			\mathop{\min}\limits_{\mathbb{C}} \sum_{i=1}^n[\hat{\textbf{Y}}_i, \textbf{Y}_i]_{\mathcal{L}_{\mathbb{C}}},
		\end{split}
	\end{equation}
	where $\mathcal{L}_{\mathbb{C}}$ denotes the loss function.
	
	\textit{\textbf{$\rhd$ CCF Module}}:
	Prior methods ~\cite{ning2021smu, RDAU-NET, vakanski2020attention, byra2020breast, CF2-Net, shahroudnejad2021thyroid, wu2020ultrasound, pan2021sgunet,yang2021efficient} generally utilize the skip connectivity operator to bridge the features from encoding and decoding blocks, which typically ignores their potential relation and may result in feature redundancy.
	With this in mind, we propose a CCF module to fuse them under the prior of correlation.
	For convenience, let $\textbf{X}_E\in \mathbb{R}^{h\times w\times c}$ and $\textbf{X}_D \in \mathbb{R}^{\frac{h}{r} \times \frac{w}{r} \times c}$ denote the low-level and high-level representations from encoding and decoding blocks, respectively.
	As illustrated in Fig.~\ref{fig_2}(e), $\textbf{X}_D$ is first up-sampled as $\hat{\textbf{X}}_D$ to obtain the same spatial resolution with $\textbf{X}_E$.
	Then, both of them are flattened into two feature matrices (i.e., $\tilde{\textbf{X}}_E \in \mathbb{R}^{hw \times c}$ and $\tilde{\textbf{X}}_D \in \mathbb{R}^{hw \times c}$). 
	After that, we calculate the relation matrix $\textbf{M} \in \mathbb{R}^{c \times c}$ by
	\begin{equation}
		\textbf{M}=\tilde{\textbf{X}}_E^T \times \tilde{\textbf{X}}_D.
		\label{eq_mr}
	\end{equation}
	Intuitively, the $p$-th row vector of $\textbf M$ (i.e., $\textbf{M}_{p,\star}$) represents the correlation between the $p$-th channel of $\textbf{X}_E$ and all channels of $\textbf{X}_D$, while the $q$-th column vector of $\textbf M$ (i.e., $\textbf{M}_{\star, q}$) records the correlation between the $q$-th channel of $\textbf{X}_D$ and all channels of $\textbf{X}_E$.
	However, $\textbf{M}_{p,\star}$ or $\textbf{M}_{\star, q}$ only reflects channel-wise relationship and lacks a global perspective to explore the relationship patterns.
	Therefore, we utilize a mapping function to learn the global relation along horizontal and vertical directions, which can be computed by
	\begin{equation}
		\bm \omega _E =\{\rho(\textbf{M}_{p,\star}^T)\}_{p=1}^c, \quad
		\bm \omega _D =\{\rho(\textbf{M}_{\star, q})\}_{q=1}^c,
		\label{eq_rv1}
	\end{equation}
	where $\bm \omega _E \in \mathbb{R}^{c \times 1}$ and $ \bm \omega _D \in \mathbb{R}^{c \times 1}$ represent two global relation vectors and $\rho(\cdot)$ denotes the mapping function that is implemented via a vertical convolution operation with the kernel size of $c \times 1$.
	Subsequently, $\bm \omega _E$ and $ \bm \omega _D$ are used to enhance the significant channels of $\textbf{X}_E$ and $\hat{\textbf{X}}_D$, respectively:
	\begin{equation}
		\bar{\textbf{X}}_E= \bm \omega_E^T \odot \textbf{X}_E,\quad
		\bar{\textbf{X}}_D= \bm \omega_D^T  \odot \hat{\textbf{X}}_D.
		\label{eq_cm1}
	\end{equation}
	Finally, $\bar{\textbf{X}}_E$ and $\bar{\textbf{X}}_D$ are concatenated and merged via a convolutional unit $f_{3 \times 3}$ whose output is considered as the output of a CCF module.
	
	\noindent{\textbf{Remark 3:}}
	\textit{Different from previous fusion strategies, CCF module conducts relation-aware fusion so as to reduce redundant information and strengthen the efficiency of feature propagation.}
	
	\subsubsection{{Coupler}}
	The coupler is composed of an auxiliary stream and an aggregation stream, as shown in Fig.~\ref{fig_2}(a). 
	The auxiliary stream aims to learn background-salient representation, while the aggregation stream is to reinforce foreground-salient representation learning with complementary background-salient information.
	\\\textit{\textbf{Auxiliary Stream}}:
	The auxiliary stream receives the concatenation of original image and background saliency map as input\footnote{We denote the concatenation of original image and background saliency map as ${\textbf E}^{b}_{0}$, namely ${\textbf E}^{b}_{0}={\textbf I} \oplus \textbf{S}^b$.}, which contains five contracting blocks $\{E^{b}_j\}_{j=1}^5$ and four expanding blocks $\{D^b_j\}_{j=1}^4$.
	For each block $E^{b}_j$ (or $D^b_j$), its architecture is similar to that of $E_j^f$ (or $D_j^f$), and substitutes a $2 \times 2$ pooling (or upsampling) layer for the DCP (or CCF) module to generate ${\textbf E}^{b}_{j}$ (or ${\textbf D}^{b}_{j}$).
	Such architecture design lies on two reasons as follows: 1) For auxiliary background path, it poses a slack requirement on elaborate detail preservation. 
	2) Plugging more DCP or CCF modules would increase the amount of parameters and is inclined to overfitting issue, which is thankless.
	\\\textit{\textbf{Aggregation Stream}}:
	The aggregation stream is constructed based on a group of aggregation units, i.e., $\{A_j\}_{j=1}^9$.
	The $j$-th aggregation unit $A_j$ takes the foreground feature $\textbf{E}_j^f$ and background feature $\textbf{E}_j^b$ as input for $j \in \{1,\cdots,5\}$, while $\textbf{D}_{j-5}^f$ and $\textbf{D}_{j-5}^b$ for $j \in \{6,\cdots,9\}$. 
	And two convolutional units $f_{1\times1}$ and $f_{3\times 3}$ are first plugged to map $\textbf{E}_j^f$ (or $\textbf{D}_{j-5}^f$) and $\textbf{E}_j^b$ (or $\textbf{D}_{j-5}^b$) into foreground and  background embeddings, respectively. 
	These two embeddings are fed into a dependency-aware reinforcement (DaR) module to generate feature $\textbf{A}_j$.
	
	\textit{\textbf{$\rhd$ DaR Module}}: Recent work has demonstrated that complementary information from background can contribute to foreground representation learning via some fusion mechanisms (e.g., channel-wise concatenation~\cite{ning2021smu} and addition strategy~\cite{zhang2021dins}), which, however, generally neglects the dependency between foreground and background representations and results in sub-optimal performance.
	To this end, the DaR module is introduced to generate the enhanced representation by modeling dependencies across foreground and background information.
	For simplicity, the foreground and background embeddings are denoted as $\textbf{F}^f \in \mathbb{R}^{h\times w\times c}$ and $\textbf{F}^b\in \mathbb{R}^{h\times w\times c}$, respectively.
	As illustrated in Fig.~\ref{fig_2}(f), ${\textbf{F}}^f$ is reshaped as $\tilde{\textbf{F}}^f$ with the size of $hw \times c$.
	Meanwhile, ${\textbf{F}}^b$ is fed into the self-reversal activation operator $\varphi$ to get $\tilde{\textbf{F}}^b$.
	Then, we compress $\tilde{\textbf{F}}^b$ into two semantic prototypes (i.e., ${\textbf{K}}\in \mathbb{R}^{1\times c}$ and ${\textbf{V}}\in \mathbb{R}^{1\times c}$) via two pyramidal branches. 
	In each branch, three convolutional units $\{f_{3\times 3}^m\}_{m=1,3,5}$ with the dilation rate of $m$ and the channel number of $c$ are concurrently utilized to extract multi-scale representations:
	\begin{equation}
		\label{eq_di}
		\tilde{\textbf{K}}^m=f_{3\times 3}^{m}(\tilde{\textbf{F}}^b), \quad \tilde{\textbf{V}}^m=f_{3\times 3}^{m}(\tilde{\textbf{F}}^b).
	\end{equation}
	The features with different scales are then concatenated alone channel direction, i.e., $\hat{\textbf{K}}=\tilde{\textbf{K}}^1\oplus \tilde{\textbf{K}}^3\oplus \tilde{\textbf{K}}^5 \in \mathbb{R}^{h\times w\times 3c}$ and $\hat{\textbf{V}}=\tilde{\textbf{V}}^1\oplus \tilde{\textbf{V}}^3\oplus \tilde{\textbf{V}}^5 \in \mathbb{R}^{h\times w\times 3c}$, which are further merged by a global average pooling and a $1\times 1$ convolutional layer to obtain $\textbf{K}$ and $\textbf{V}$.
	These two compact prototypes from the background embedding are used to explore and broadcast foreground complementary information, accompanied with modeling their dependeny on foreground-salient representation.
	Specifically, the dependency is computed by exploiting $\tilde{\textbf{F}}^f$ to query and activate the foreground-related features of ${\textbf{K}}$, which is then matched with ${\textbf{V}}$ to obtain the foreground complementary representation $\tilde{\textbf{F}} \in \mathbb{R}^{hw \times c}$:
	\begin{equation}
		\label{eq_qkv}
		\tilde{\textbf{F}}=\psi(\tilde{\textbf{F}}^f\times{\textbf{K}}^T)\times{\textbf{V}}.
	\end{equation}
	Subsequently, $\tilde{\textbf{F}}$ is reshaped into the size of $h \times w \times c$, followed by a convolutional unit $f_{1 \times 1}$ to generate ${\textbf{F}} \in \mathbb{R}^{h \times w \times 2c}$.
	Finally, ${\textbf{F}}$ is added to the concatenation of ${\textbf{F}}^f$ and $\tilde{\textbf{F}}^b$ to reinforce foreground representation, namely ${\textbf{F}} + \{{\textbf{F}}^f \oplus\tilde{\textbf{F}}^b\}$, which is regarded as the output of a DaR module.
	
	\noindent{\textbf{Remark 4:}}
	\textit{Coupler, as a novel component, is introduced to learn background-salient representation and explore its dependency on foreground, upon which complementary information derived from background is incorporated to reinforce foreground-salient representation learning.}
	
	\subsection{Optimization Function}
	We design a hybrid loss to effectively train the proposed DC-Net, in which the decomposition subnet $\mathbb{D}$ and coupling subnet $\mathbb{C}$ are optimized jointly. 
	According to Eq.(\ref{eq_att4}) and Eq.(\ref{eq_Lc}), the total loss $\mathcal{L}$ can be formulated as
	\begin{equation}
		\label{eq_all}
		\mathcal{L} = {\lambda_1}{\mathcal{L}_{\mathbb{D}_f}}+{\lambda_2}{\mathcal{L}_{\mathbb{D}_b}}+{\lambda_3}{\mathcal{L}_\mathbb{C}},
	\end{equation}
	where $\lambda_1$, $\lambda_2$, and $\lambda_3$ are weight coefficients. 
	Insufficient attention to hard samples encountered during the training stage is still a challenge for accurate ultrasound lesion segmentation, since these samples (i.e., pixels around blur edge and appearance) generally occupy a smaller proportion compared to the easy ones.
	Although focal loss~\cite{li2021multi, zhu2019anatomynet} has shown its efficacy on increasing the network's attention to hard samples by reshaping CE loss such that it down-weights the loss assigned to well-segmented samples, it also reduces the loss amplitude of all samples and suppresses relatively low-confidence samples (e.g., $p=.4\sim.75$) especially when $\gamma$ increases.
	It is expected that, in the scenario of easy samples being down-weighted, the loss margin and gradient of hard samples can still be preserved and even increased to achieve continuous attention, while the range of suppression can be narrowed so as to guarantee the network to focus on low-confidence samples.
	Therefore, we introduce a harmonic loss function for ultrasound lesion segmentation.
	\\$\textbf{Definition 2 (Harmonic Loss Function $\mathcal{L}_h$).}$ 
	\textit{The harmonic loss function is defined as
		\begin{equation}
			\label{eq_fc1}
			\begin{aligned}
				\mathcal{L}_h\left( p_t \right) =-\frac{(1+\sigma )\left( 1-p_t \right) ^{\gamma -1}}{\sigma +\left( 1-p_t \right) ^{\gamma}}\log \left( p_t \right),
			\end{aligned}
		\end{equation}
		where $\sigma \geq 0$ and $\gamma$ are harmonic parameters, $p_t$ equals to $p$ for $y=1$ and $1-p$ for $y=0$, and $p$ and $y$ denote the estimated probability and label of a pixel, respectively.
		We can derive several interesting properties as follows:}
	\\\textbf{P-1.} \textit{The function is non-negative, continuous and differentiable.}
	\\\textbf{P-2.} \textit{$\mathcal{L}_h\left( p_t \right) =-\left(1-p_t \right)^{-1}\log \left(p_t \right)$, if $\sigma \rightarrow 0$.}
	\\\textbf{P-3.} \textit{$\mathcal{L}_h\left( p_t \right) =-\left(1-p_t \right)^{\gamma -1}\log \left(p_t \right)$, if $\sigma\rightarrow+\infty$.}
	\\\textbf{P-4.} \textit{$\mathcal{L}_h(p_t) =-\left(1+\sigma)(1-p_t \right)^{-1}\log \left(p_t \right)$, if $\sigma \ll (1-p_t)^{\gamma}$.}
	\\\textbf{P-5.} \textit{$\mathcal{L}_h(p_t) =-\frac{(1+\sigma)}{\sigma}(1-p_t)^{\gamma-1}\log (p_t)$, if $\sigma \gg (1-p_t)^{\gamma}$.}
	
	Obviously, the \textbf{P-1} holds, thus it is suitable as a loss function.
	The \textbf{P-2} suggests that $\mathcal{L}_h$ degenerates into CE loss $\mathcal{L}_{c}$ with the modulating factor $(1-p_t)^{-1}$ when $\sigma$ approaches zero.
	Inspired by~\cite{leng2021polyloss}, we rewrite $\mathcal{L}_{c}$ and $\mathcal{L}_h|_{\sigma\rightarrow0}$ in a Taylor expansion form:
	\begin{small}
		\begin{equation}
			\label{eq_fc2}
			\mathcal{L}_{c}\left( p_t \right)=-\log \left( p_t \right) =\sum_{j=1}^{\infty}\frac{1}{j}\left( 1-p_t \right) ^j,
		\end{equation}
	\end{small}%
	\begin{small}
		\begin{equation}
			\label{eq_fc3}
			\mathcal{L}_h|_{\sigma\rightarrow0}\left( p_t \right)=-(1-p_t)^{-1}\log \left( p_t \right) =\sum_{j=1}^{\infty}\frac{1}{j}\left( 1-p_t \right) ^{j-1}.
		\end{equation}
	\end{small}%
	Accordingly, their gradients can be computed by summating series polynomials as follows:
	\begin{small}
		\begin{equation}
			\label{eq_fc4}
			-\frac{\mathrm{d}\mathcal{L}_{c}}{\mathrm{d}p_t}=\sum_{j=1}^{\infty}{\left( 1-p_t \right) ^{j-1}}\\=1+\left( 1-p_t \right) +\left( 1-p_t \right) ^2+\cdot \cdot \cdot,
		\end{equation}
	\end{small}%
	\begin{equation}
		\label{eq_fc5}
		\begin{aligned}
			-\frac{\mathrm{d}\mathcal{L}_h|_{\sigma\rightarrow0}}{\mathrm{d}p_t}=&\sum_{j=1}^{\infty}{\frac{j-1}{j}\left( 1-p_t \right) ^{j-2}}\\=&\frac{1}{2}+\frac{2}{3}\left( 1-p_t \right) +\frac{3}{4}\left( 1-p_t \right) ^2+\cdot \cdot \cdot.
		\end{aligned}
	\end{equation}
	Intuitively, $\mathcal{L}_h|_{\sigma\rightarrow0}$ provides a smaller constant gradient than $\mathcal{L}_c$, which can prevent the model from emphasizing the majority class (i.e., easy samples)~\cite{leng2021polyloss}.
	Thus, it can be regarded as an improved CE loss.
	Similarly, \textbf{P-3} indicates $\mathcal{L}_h$ is an equivalent variant of focal loss $\mathcal{L}_f$ if $\sigma$ approaches the infinite, and their gradients can be calculated by
	\begin{equation}
		\label{eq_fc6}
		-\frac{\mathrm{d}\mathcal{L}_f}{\mathrm{d}p_t}=\sum_{j=1}^{\infty}{(1+\frac{\gamma}{j})\left( 1-p_t \right) ^{j+\gamma-1}},
	\end{equation}
	\begin{equation}
		\label{eq_fc7}
		-\frac{\mathrm{d}\mathcal{L}_h|_{\sigma\rightarrow+\infty}}{\mathrm{d}p_t}=\sum_{j=1}^{\infty}{(1+\frac{\gamma-1}{j})\left( 1-p_t \right) ^{j+\gamma-2}}.
	\end{equation}
	It is apparent that $\mathcal{L}_h|_{\sigma\rightarrow+\infty}$ is equivalent to shifting all the polynomial coefficients of $\mathcal{L}_f$ by 1.
	As to \textbf{P-4} and \textbf{P-5}, they suggest that $\mathcal{L}_h$ additionally introduces the amplification factors $1+\sigma$ and $\frac{1+\sigma}{\sigma}$ for $\mathcal{L}_h|_{\sigma\rightarrow0}$ and $\mathcal{L}_h|_{\sigma\rightarrow+\infty}$, respectively, when the conditions of $\sigma \ll (1-p_t)^{\gamma}$ and $\sigma \gg (1-p_t)^{\gamma}$ are satisfied.
	\begin{figure}[!t]
		\centering
		\includegraphics[width=1\linewidth]{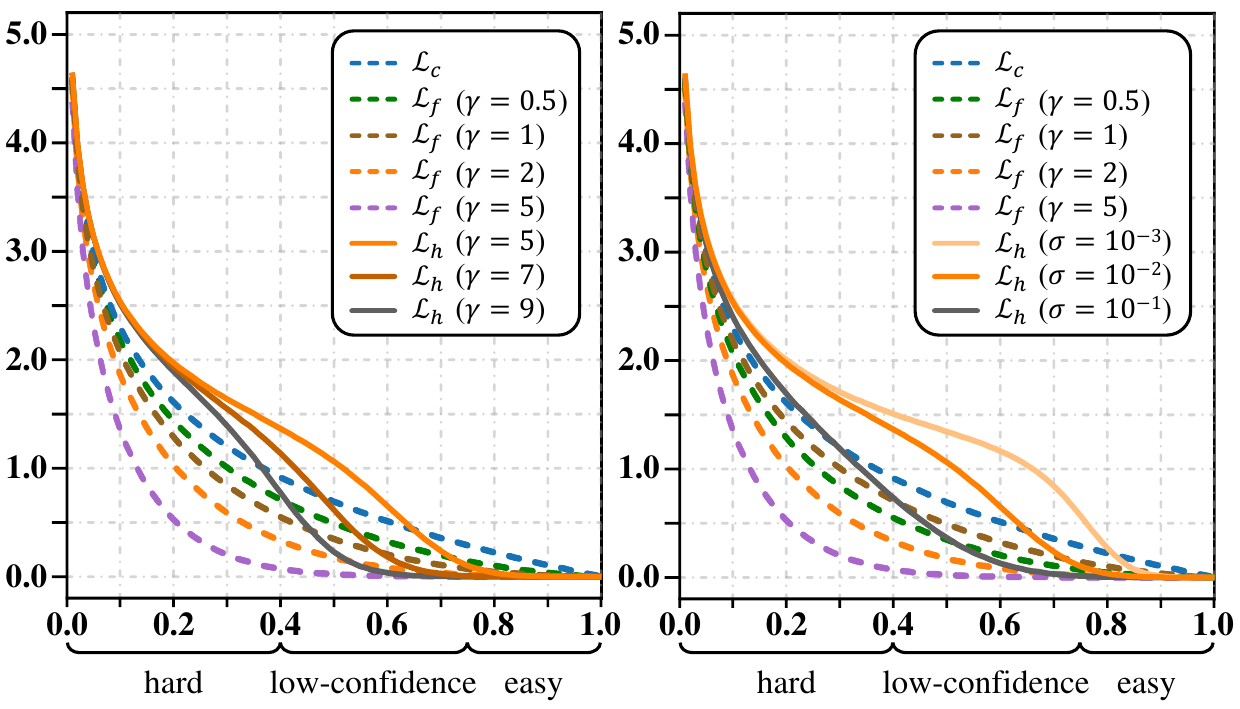}
		\caption{Illustration and comparison of cross-entropy loss ($\mathcal{L}_c$), focal loss ($\mathcal{L}_f$ with $\gamma =2$ and $\gamma=5$) and harmonic loss ($\mathcal{L}_h$ with $\gamma =5, \gamma=7, \gamma=9$ and $\sigma=10^{-3}, \sigma=10^{-2}, \sigma=10^{-1}$). 
			For facilitating comparison, $\mathcal{L}_c$ and $\mathcal{L}_h$ are served as the control group and showed in each subFigure.}
		\label{fig_3}
		\vspace{-5mm}
	\end{figure}
	Furthermore, we integrative different settings of $\sigma$ and $\gamma$ for $\mathcal{L}_h$, and provide an intuitive comparison among $\mathcal{L}_h$, $\mathcal{L}_f$ and $\mathcal{L}_c$ in Fig.~\ref{fig_3}.
	Several key points can be observed as follows:
	1) Compared with $\mathcal{L}_c$, $\mathcal{L}_h$ down-weights easy samples as $\mathcal{L}_f$ does;
	2) $\mathcal{L}_h$ almost preserves and even elevates the loss margin and gradient of hard samples;
	3) $\mathcal{L}_h$ narrows down the suppression range to prevent low-confidence samples from being overwhelmed, and the range would shrinkage as either $\gamma$ or $\sigma$ decreases.
	Additionally, we introduce a weighting factor $\alpha_t \in [0,1]$ to alleviate imbalance between foreground and background, and the Eq.(\ref{eq_fc1}) can be rewritten as
	\begin{equation}
		\label{eq_fc9}
		\mathcal{L}_{h}(p_t) =-\alpha_t{\frac{(1+\sigma)(1-p_t)^{\gamma-1}}{\sigma+(1-p_t)^{\gamma}}} log(p_t),
	\end{equation}
	where $\alpha_t=\alpha$ if $y=1$, otherwise $\alpha_t=1-\alpha$ ($\alpha$ is a hyper-parameter).
	
	Considering that compound loss function would help obtain a robust and accurate model for medical image segmentation~\cite{MA2021102035lossodyssey}, we design $\mathcal{L}_\mathbb{C}$ by incorporating $\mathcal{L}_{h}$ with dice similarity coefficient (Dice) loss $\mathcal{L}_{d}$ \cite{milletari2016v}:
	\begin{equation}
		\label{eq_fcd}
		\begin{split}
			\mathcal{L}_\mathbb{C} &=\mathcal{L}_d + {\lambda_4} \mathcal{L}_h,
		\end{split}
	\end{equation}
	where ${\lambda_4}$ is a weight coefficient.
	Since decomposition subnet aims to preliminarily disentangles the original image into saliency maps and it does not require accurate estimation, we thus set ${\mathcal{L}_{\mathbb{D}_f}}=\mathcal{L}_d$ and ${\mathcal{L}_{\mathbb{D}_b}}=\mathcal{L}_d$ to reduce the computational cost.
	
	\noindent{\textbf{Remark 5:}}
	\textit{Compared with focal loss, our harmonic loss can not only down-weights easy samples, but also preserves the loss margin and gradient of hard ones. More importantly, it can adjust the suppression range to prevent low-confidence samples from being overwhelmed via the harmonic parameters $\sigma$ or $\gamma$.}

	\section{Experiments and Results}
	In this section, we first introduce datasets and experimental settings.
	Then, we present experimental results, including comparison analysis as well as ablation analysis.
	
	\subsection{Datasets and Experimental Settings}
	We evaluated our proposed method on two different ultrasound lesion segmentation tasks, i.e., breast ultrasound (BUS) segmentation and thyroid ultrasound (TUS) segmentation. 
	Five multi-site BUS datasets that contains 2862 images and a TUS dataset which enrolls 646 images were used to develop and assess the proposed method.
	Brief sample distribution of BUS and TUS datasets is summarized in TABLE~\ref{tableD}.
	\\\textbf{BUS}.
	S1: Dataset1 collects 1812 BUS images offered by the SonoSkills and Hitachi Medical Systems Europe\footnote{https://www.ultrasoundcases.info/};
	S2: Dataset2 has 200 BUS images acquired by the Ultrasonix SonixTouch Research ultrasound scanner with a L14-5/38 linear array transducer~\cite{piotrzkowska2017open};
	S3: Dataset3 contains 163 BUS images provided by the UDIAT Diagnostic Center with a Siemens ACUSON Sequoia C512 system  and a 17L5 HD linear array transducer ($8.5$ MHZ)~\cite{yap2017automated}.
	S4: Dataset4 enrolls 506 BUS images derived from the Baheya Hospital with a LOGIQ E9 ultrasound system or LOGIQ E9 Agile ultrasound system~\cite{al2020dataset}.
	S5: Dataset5 includes 181 BUS images obtained by a Philips iU22 ultrasound scanner at Thammasat University Hospital~\cite{rodtook2018automatic}.  
	The ground-truths of all datasets have been provided by the organizers except for Dataset3.
	Thus, the annotations of Dataset3 have been manually delineated by a radiologist with two years of clinical experience and confirmed by another radiologist with seven years of clinical experience.
	Also, they have reviewed the annotations of other datasets and removed partial poor-quality images that may result in inconsistent suggestions.
	\\\textbf{TUS}. 
	The dataset recruits 646 B-mode TUS images and their ground-truths, which is provided by the IDIME Ultrasound Department in Colombia~\cite{pedraza2015open}.
	The radiologists have been also invited to review and confirm image's quality as well as annotation.
	
	\begin{table}[t]
		\setlength{\abovecaptionskip}{0.cm}
		\renewcommand{\arraystretch}{1.3}
		\caption{{Brief sample distribution of two datasets.}}
		\label{tableD}
		\centering
		\small
		\setlength{\tabcolsep}{0.8mm}{
			\begin{tabular}{cccccc|c}
				\hline
				\multicolumn{1}{c}{} &  \multicolumn{5}{c|}{{BUS}}  & \multicolumn{1}{c}{{TUS}} \\ 
				\hline
				\multicolumn{1}{c}{}   & \multicolumn{1}{c}{Dataset1} & \multicolumn{1}{c}{Dataset2} & \multicolumn{1}{c}{Dataset3} & \multicolumn{1}{c}{{Dataset4}} & \multicolumn{1}{c|}{{Dataset5}} & \multicolumn{1}{c}{{--}} \\ 
				\hline
				Benign    & 818  & {96}  & 110     & 383    & {121}       & {66} \\
				Malignant & 907  & {104} & 53      & 123    & {60}        & {380} \\
				Unknow    & 87  & {-}   & -       & -      & {-}         & {200} \\
				\hline
				Total     & 1812 & {200} & 163     & 506    & {181}       & 646  \\
				\hline      
		\end{tabular}}
		\vspace{-5mm}
	\end{table}
	
	For BUS segmentation task, we trained the model on two sites (i.e., S1 and S2~\cite{piotrzkowska2017open}, 2012 images in total) via inner 5-fold cross-validation strategy and tested on three sites (i.e., S3~\cite{yap2017automated}, S4~\cite{al2020dataset} and S5~\cite{rodtook2018automatic}, 850 images in total) to assess model's generalization capability on multiple sites.
	As for TUS segmentation task, we utilized 5-fold cross-validation strategy to evaluate the proposed method in view of the limited sample size~\cite{pedraza2015open}.
	All images were resized into a fixed size of $256\times256$ and the intensity values in each image were normalized into the range of [0, 1] by Min-Max normalization.
	The performance of model was assessed in terms of dice similarity coefficient (Dice), mean intersection over union (m-IoU), precision and 95\% hausdorff distance (95HD).
	The parameters $\lambda_1$, $\lambda_2$, $\lambda_3$, $\lambda_4$, $\sigma$, $\gamma$ and $\alpha$ were experimentally set to 0.5, 0.5, 1, 10, 0.001, 5 and 0.25, respectively.
	Notably, the influence of key parameters on model's performance has been discussed in the section of \textbf{Discussion}.
	We initialized all parameters with "he\_normal" and adopted Adam solver with the momentum of 0.9 for network training.
	The batch size, learning rate and epoch number were set to 4, $3\times 10^{-4}$ and 200, respectively.
	All intensive calculations were offloaded to two 12 GB NVIDIA Pascal Titan X GPU.

	\subsection{Comparison with the State-of-the-art Methods}
	We first compared the proposed method with the state-of-the-art deep learning approaches in ultrasound lesion segmentation.
	The competing methods include: 
	1) Classical medical image segmentation models:
	DeepLab V3+~\cite{chen2018encoder}, U-Net~\cite{ronneberger2015unet}, U-Net++~\cite{Unet++}, and AU-Net~\cite{Oktay};
	2) Breast lesion segmentation models:
	RDAU-Net~\cite{RDAU-NET}, AE-Net~\cite{vakanski2020attention}, ConvEDNet~\cite{Boundary}, {SK-Net~\cite{byra2020breast}}, CF$^2$-Net~\cite{CF2-Net}, MNFE-Net~\cite{liu2023multiscale}, and CSwin-PNet~\cite{yang2023cswin};
	3) Thyroid nodule segmentation models:
	TNSCUI-R1~\cite{wang2020automatic}, TNSCUI-R2~\cite{chen2021lrthr}, resDU-Net~\cite{shahroudnejad2021thyroid}, JU-Net~\cite{wu2020ultrasound}, and SGU-Net~\cite{pan2021sgunet}, FDE-Net~\cite{chen2023fde}, and BPAT-UNet~\cite{bi2023bpat}.
	As not all comparison methods have been evaluated on the same datasets, we reimplemented them based on the details and source codes provided by their work.
	Table~\ref{table1} shows the experimental results.

	\begin{table*}[!t]
		\setlength{\abovecaptionskip}{0.cm}
		\renewcommand{\arraystretch}{1.3}
		\caption{Segmentation results (mean ${\pm}$ standard deviation) of all competing methods on BUS and TUS datasets. And the \textit{p}-value of statistical test is also provided.}
		\label{table1}
		\centering
		\scriptsize
		\setlength{\tabcolsep}{.8mm}{
			\begin{tabular}{c|c|c|c|c|c|c|c|c|c|c|c|c}
				\hline
				Methods      	& DeepLab V3+  	&U-Net        	&U-Net++      	&AU-Net       	&RDAU-Net     	&AE-Net       	&ConvEDNet    	&CF$^2$-Net   	&SK-Net       	&MNFE-Net	&CSwin-PNet	&DC-Net  \\ \hline    
				Dice           	 &63.13$\pm$1.63 	 &63.85$\pm$3.04 	 &65.82$\pm$1.60 	 &67.37$\pm$1.54 	 &61.94$\pm$0.96 	 &66.88$\pm$1.78 	 &68.90$\pm$0.72 	 &69.04$\pm$1.94 	 &70.33$\pm$1.22 	& 66.70$\pm$0.61	 &70.99$\pm$1.75 	 &\textbf{73.96$\pm$0.36}  \\
				m-IoU          	& 52.11$\pm$1.91 	& 53.29$\pm$3.25 	
				& 55.22$\pm$1.58 	& 57.31$\pm$1.54 	& 51.90$\pm$0.94 &56.69$\pm$1.75 	& 58.00$\pm$0.65 	& 59.62$\pm$2.31 &60.87$\pm$1.45 	& 55.88$\pm$0.77	 &61.16$\pm$2.48 	 &\textbf{64.26$\pm$0.48}   \\
				Precision      	& 61.55$\pm$3.34 	& 60.90$\pm$4.23 	
				& 63.23$\pm$2.44 	& 65.07$\pm$2.16 	& 62.07$\pm$1.63 	
				& 64.43$\pm$2.57 	& 66.87$\pm$1.06 	& 67.66$\pm$3.01 	
				& 69.81$\pm$2.41 	& 67.00$\pm$2.64	& 70.16$\pm$1.96	 &\textbf{75.82$\pm$1.16}   \\
				95HD          	& 6.45$\pm$0.19 	& 6.67$\pm$0.44  &6.35$\pm$0.22 	& 6.24$\pm$0.14 	& 6.38$\pm$0.11 	 &6.22$\pm$0.18 	& 5.92$\pm$0.07 	& 5.90$\pm$0.23  &5.88$\pm$0.07 	& 6.10$\pm$0.16 	& 5.75$\pm$0.07 	 &\textbf{5.44$\pm$0.10} \\
				\textit{p}-value 	& 7.59e-05 	& 2.82e-03  	& 7.68e-04 	
				& 5.69e-04  	& 2.90e-06  	& 9.05e-04   	& 1.78e-05  
				& 6.14e-03 	& 2.94e-03    	&1.78e-05	&3.10e-02	& ---  \\\hline
				Methods	        &DeepLab V3+ 	&U-Net       	&U-Net++     	&AU-Net      	&TNSCUI-R1   	&resDU-Net   	&JU-Net      	&SGU-Net     	&TNSCUI-R2   	&FDE-Net	    & BPAT-UNet	 
				& DC-Net    \\  \hline
				Dice            &62.67$\pm$3.81 	 &71.65$\pm$2.22 	 &67.35$\pm$1.94            	& 73.40$\pm$1.89 	 &62.81$\pm$2.66 	 &64.36$\pm$2.01 	 &69.79$\pm$0.76          	 &72.07$\pm$0.35 	 &72.13$\pm$2.37          	 &69.55$\pm$1.11            	 &73.40$\pm$0.95            	 &\textbf{76.25$\pm$1.89} \\
				m-IoU           	 &48.63$\pm$4.42 	 &59.13$\pm$2.52 	 &53.93$\pm$2.19 	 &61.32$\pm$2.06 	 &49.03$\pm$2.89 	 &50.11$\pm$2.03 	 &56.80$\pm$0.57   	 &59.33$\pm$0.60    	 &61.25$\pm$2.69          	 &56.61$\pm$1.83	& 61.07$\pm$1.26 	 &\textbf{65.28$\pm$2.24} \\
				Precision         	 &64.16$\pm$6.52 	& 73.25$\pm$2.45 	 &67.55$\pm$1.95 	 &76.99$\pm$4.08 	 &64.71$\pm$4.23 	 &64.29$\pm$2.84 	 &72.52$\pm$1.68    	 &75.46$\pm$2.42 	 &72.06$\pm$1.64          	 &73.43$\pm$2.74	 &74.21$\pm$5.54          	 &\textbf{79.35$\pm$1.92} \\
				95HD 	 &7.64$\pm$0.43	 &6.81$\pm$0.27	 &7.26$\pm$0.17	 &6.56$\pm$0.23	 &7.28$\pm$0.24	 &7.41$\pm$0.21	 &6.83$\pm$0.11	 &6.86$\pm$0.10	 &9.46$\pm$0.06	 &6.81$\pm$0.27             	 &6.60$\pm$0.12             	 &\textbf{6.22$\pm$0.26}\\
				\textit{p}-value     & 4.80e-03	&6.90e-05	&2.02e-04	&4.17e-03	&5.44e-04	&2.51e-04	&4.89e-04	&4.32e-03	&4.12e-02	&5.42e-03	&3.81e-02	 &--- \\\hline        
		\end{tabular}}
		\vspace{-5mm}
	\end{table*}

	We have several key observations. 
	\textit{First}, all U-Net variants outperform DeepLab V3+ on both ultrasound lesion segmentation tasks, which might benefit from the skip connectivity operator in U-Net variants.
	\textit{Second}, AU-Net obtains remarkable performance gain than most U-Net variants, especially on TUS segmentation task. 
	It further indicates that attention mechanism can help boost model's performance for lesion segmentation~\cite{Oktay}.
	\textit{Third}, multi-scale strategy is competitive across model's capability of learning features for lesions in complex-scenario images.
	For example, SK-Net produces higher segmentation accuracy by expanding receptive field to extract multi-scale contexts via the dilated convolution, while TNSCUI-R2 adopts the ensemble refinement mechanism to iteratively optimize the network with multi-resolution images.
	\textit{Fourth}, ConvEDNet displays pretty promising in BUS segmentation and is competitive with CF$^2$-Net.
	It is likely that the utilization of edge constraint advances edge recognition in low-contrast and blurry-boundary images.
	\begin{figure*}[t]
		\centering
		\includegraphics[width=1\linewidth]{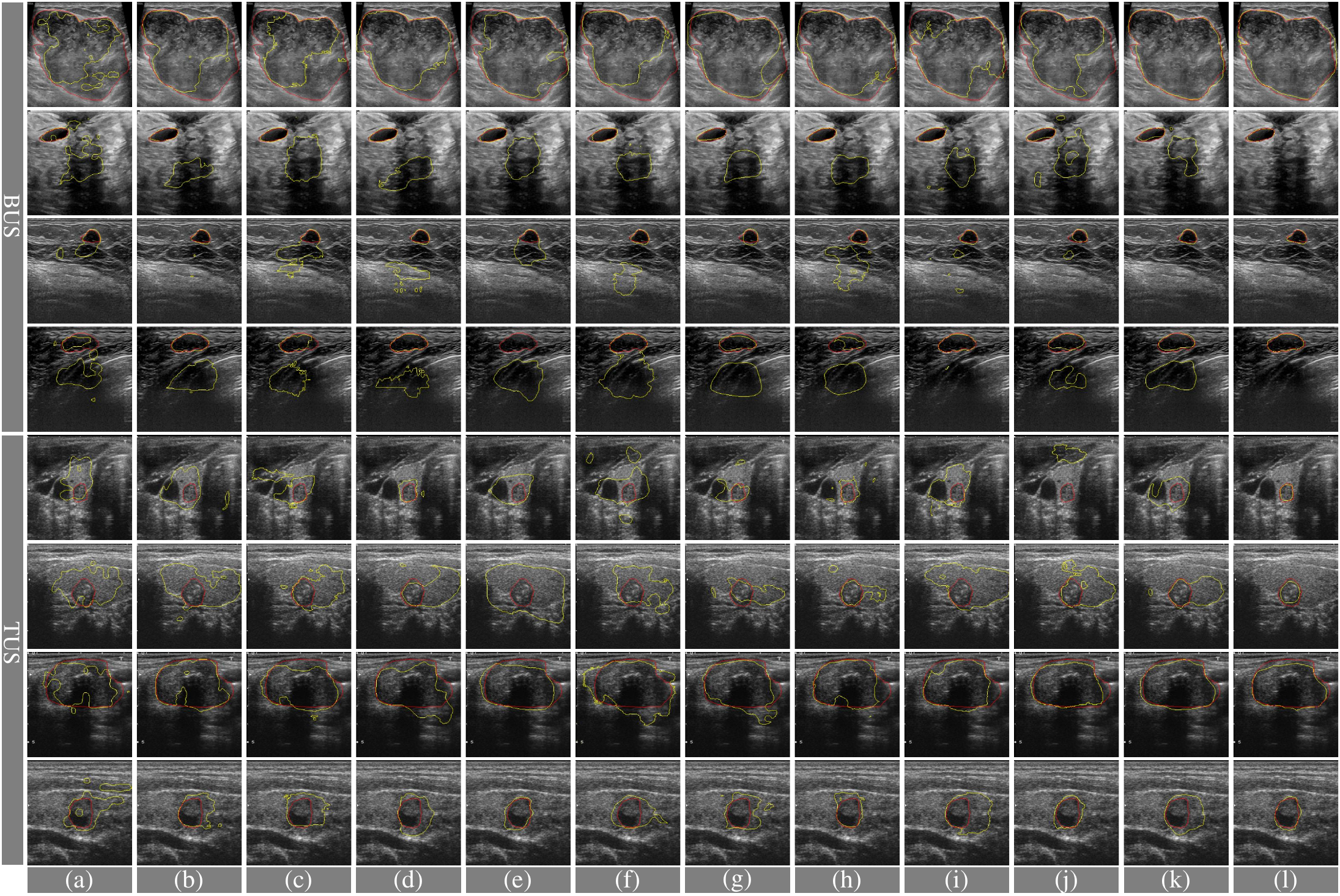}
		\caption{Visualization results of all competing methods on some representative cases from BUS and TUS datasets, including: (a) DeepLab V3+, (b) U-Net, (c) U-Net++, (d) AU-Net, (e) RDAU-Net/TNSCUI-R1, (f) AE-Net/resDU-Net, (g) ConvEDNet/JU-Net, (h) CF$^2$-Net/SGU-Net, (i) SK-Net/TNSCUI-R2, (j) MNFE-Net/FDE-Net, (k) CSwin-PNet/BPAT-UNet, and (l) DC-Net. The red and yellow curves represent the ground truth and the segmentation results of different compared methods, respectively.}
		\label{figS1}
		\vspace{-7mm}
	\end{figure*}
	\textit{Fifth}, well pre-trained backbones may help to improve model performance.
	For example, CSwin-PNet and BPAT-UNet introduced pre-trained transformer blocks to improve the network's capability of extracting long-range dependency and achieved better segmentation results.
	\textit{Finally}, the proposed DC-Net which we find to reach the noticeably higher performance in terms of all evaluation metrics when compared with other methods.
	Several potential advantages exist in the proposed method:
	1) It integrates saliency map generation and lesion segmentation into a unified framework, and introduces a coupler to reinforce foreground-salient representation learning by incorporating complementary information derived from background-salient cues.
	2) It performs regional feature aggregation via DCP operator to alleviate the issue of information dropout by adaptively preserving local important contextual details and enlarging receptive field. 
	3) It strengthens the connection of information from encoding and decoding streams (via CCF) to efficiently conduct relation-aware representation fusion of low-level and high-level features.
	4) The harmonic loss encourages the network to focus more attention on hard and low-confidence samples during the training stage, which helps improve generalization ability.
	In Fig.~\ref{figS1}, we also provide the visualization results of all competing methods on some representative cases for further comparison.
	From Fig.~\ref{figS1}, we can find that DC-Net can more accurately locate the lesion than others, which mainly credits to the utilization of the cues provided by saliency maps.
	Also, DC-Net can work well in the scenario where lesion region shares similar intensity with surrounding tissues, especially on the aspect of edge's integrity.

	\begin{figure}[!t]
		\centering
		\includegraphics[width=1\linewidth]{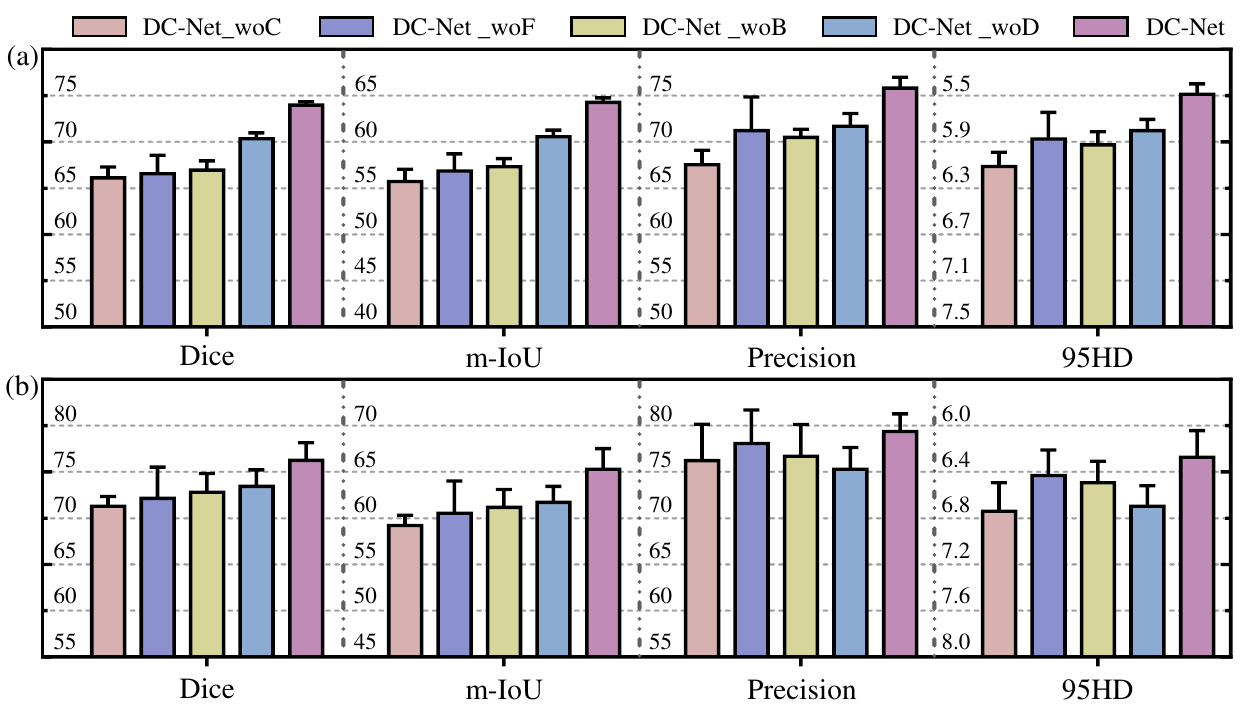}
		\caption{Bar plots of four evaluation metrics (mean $\pm$ standard deviation) achieved by DC-Net and its variants (for validating the efficacy of saliency map) on (a) BUS and (b) TUS datasets.}
		\label{fig_4}
		\vspace{-5mm}
	\end{figure}
	
	\begin{figure*}[!t]
		\centering
		\includegraphics[width=1\linewidth]{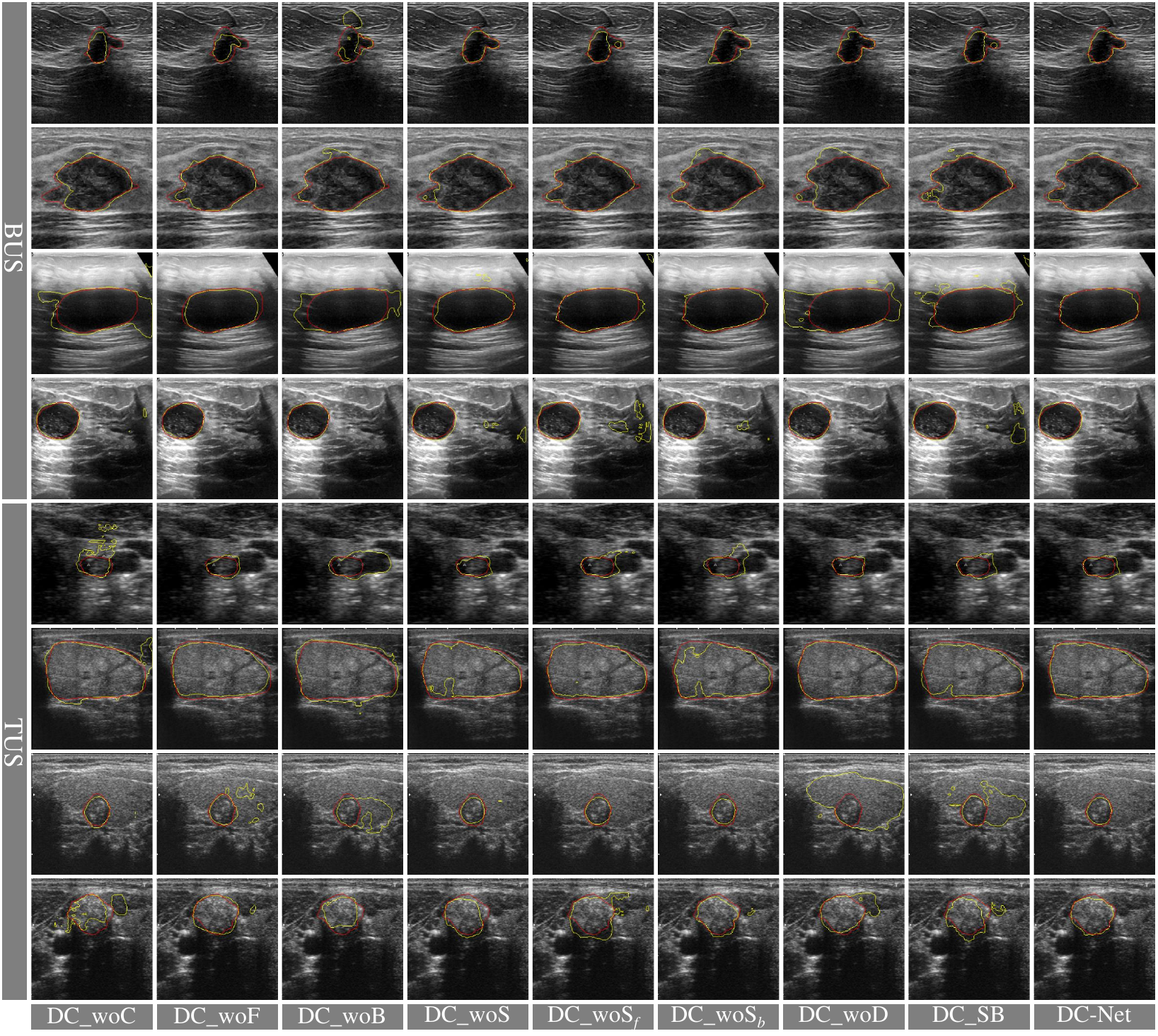}
		\caption{Visualization results of the variants of saliency map on some representative cases from BUS and TUS datasets. The red and yellow curves represent the ground truth and the segmentation results of different variants, respectively.}
		\label{figS2}
		\vspace{-4mm}
	\end{figure*}

	\subsection{Efficacy of Saliency Map and DaR Module}
\begin{figure}[t]
		\centering
		\includegraphics[width=1\linewidth]{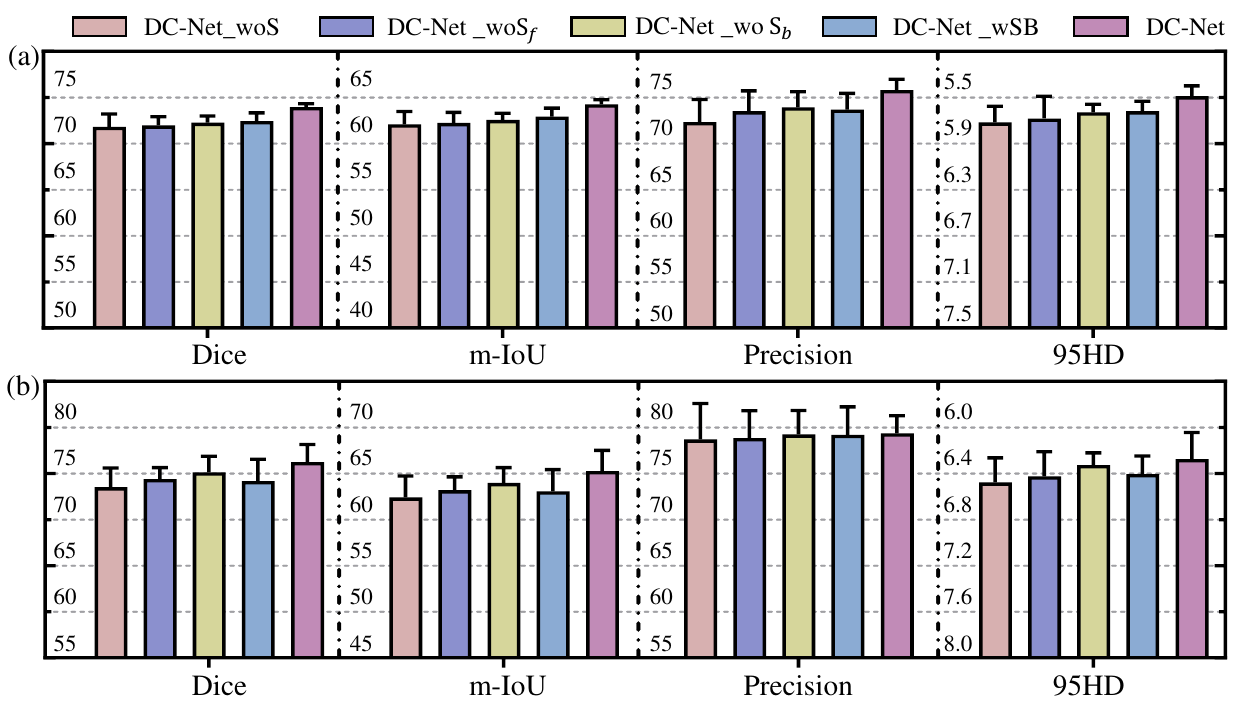}
		\caption{Bar plots of four evaluation metrics (mean $\pm$ standard deviation) achieved by DC-Net and its variants (for validating the efficacy of supervision strategy) on (a) BUS and (b) TUS datasets.}
		\label{fig_5}
		\vspace{-2mm}
	\end{figure}

\begin{figure}[!t]
		\centering
		\includegraphics[width=1\linewidth]{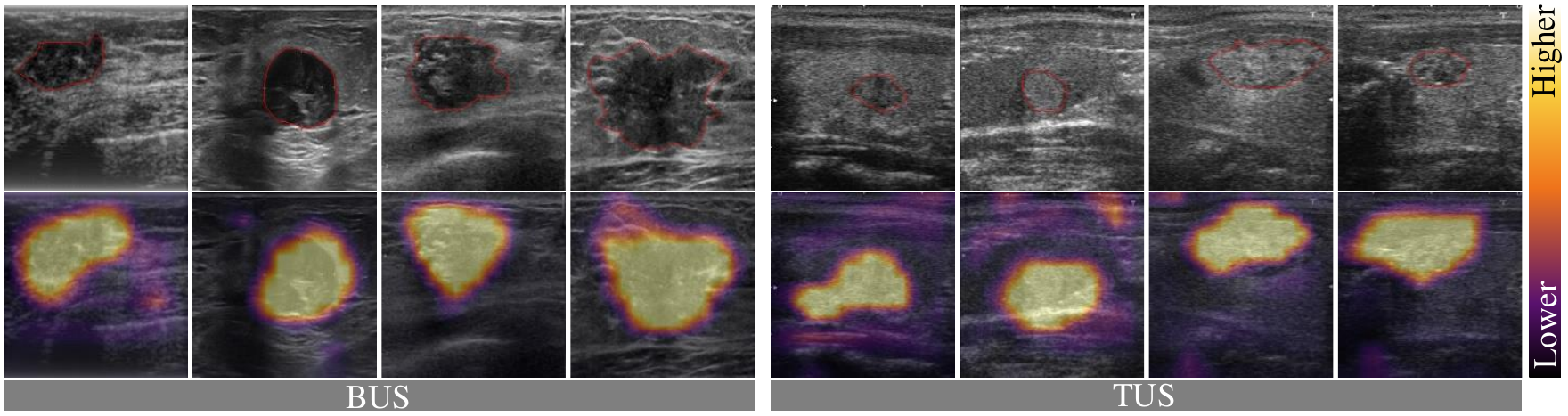}
		\caption{Visualization results of attention map (i.e., $\psi(\tilde{\textbf{F}}^f\times{\textbf{K}}^T)$) in the DaR module on some representative cases from BUS and TUS datasets. Red line on the original image represents the ground truth.}
		\label{fig_dar_vis}
		\vspace{-4mm}
	\end{figure}

	The saliency map is one of core components of DC-Net and DaR module helps couple the saliency maps of background and foreground.
	In this part, we aim to validate their efficacy by developing and evaluating various variants.
	For convenience, we name these variants as follows.
	1) DC-Net\_woB (DC-Net\_woF): It removes coupler and only uses the concatenation of original image and foreground (background) saliency map as the input of coupling subnet; 
	2) DC-Net\_woC: It removes both decomposition subnet and coupler, and only takes the original image as the input of coupling subnet.
	3) DC-Net\_woD: It abandons DaR module and links foreground-salient and background-salient representations by channel-wise concatenation operation.
	The experimental results are shown in Fig.~\ref{fig_4}.
	The DC-Net\_woC is inferior to others on two segmentation tasks, which verifies the conclusion that saliency map is a significant step towards complex-scenario ultrasound lesion segmentation~\cite{xu2016deep, wang2018deepigeos, ning2021smu}.
	And DC-Net significantly overpasses both DC-Net\_woB and DC-Net\_woF, owing to the coupler that can extract discriminative information derived from background to complement foreground-salient representation, especially when background contains richer texture patterns than foreground.
	Also, the DC-Net\_woD is found to suffer from a profound performance degradation, and this degradation suggests DaR module can better help reinforce foreground-salient representation compared to simple concatenation operation that may introduce more irrelevant information.
	Fig.~\ref{fig_dar_vis} shows the attention map (i.e., $\psi(\tilde{\textbf{F}}^f\times{\textbf{K}}^T)$) in the DaR module on some representative cases from BUS and TUS datasets. 
	From the Fig.~\ref{fig_dar_vis}, we can find that the attention map mainly concentrates on the foreground (i.e., lesion region), which can be utilized to mine foreground-dependent complementary information derived from background and reinforce the foreground-salient representation learning.
	To further investigate the effectiveness of supervision strategy for decomposition subnet, we implement DC-Net with only the minor modification of using different supervision schemes.
	1) DC-Net\_woS$_f$ (DC-Net\_woS$_b$): It discards the supervision of foreground (background) mask on saliency map generation;
	2) DC-Net\_woS: It trains decomposition subnet in an unsupervised manner;
	3) DC-Net\_wSB: It replaces the precise mask with bounding box to optimize saliency map generation.
	In Fig.~\ref{fig_5}, as we expected, DC-Net\_woS underperforms other variants, since supervision signal can restrain the uncertainty of parameters and accelerate the convergence of model.
	Additionally, DC-Net surpasses DC-Net\_woS$_f$ and DC-Net\_woS$_b$, which indicates that the cooperation of foreground and background masks can help to efficiently disentangle the original image into foreground and background saliency maps and strengthen model's performance.
	Interestingly, DC-Net\_wSB is also likely to achieve promising performance, implying proxy annotation may be an alternative choice for training saliency map generation.
	Fig.~\ref{figS2} displays some representative cases from BUS and TUS datasets for further comparison.
	From Fig.~\ref{figS2}, we can see that the model can show better segmentation performance by simultaneously learning foreground and background representation under the assistance of foreground and background saliency maps. 
	
	\subsection{Efficacy of DCP Operator}
\begin{figure}[!t]
		\centering
		\includegraphics[width=1\linewidth]{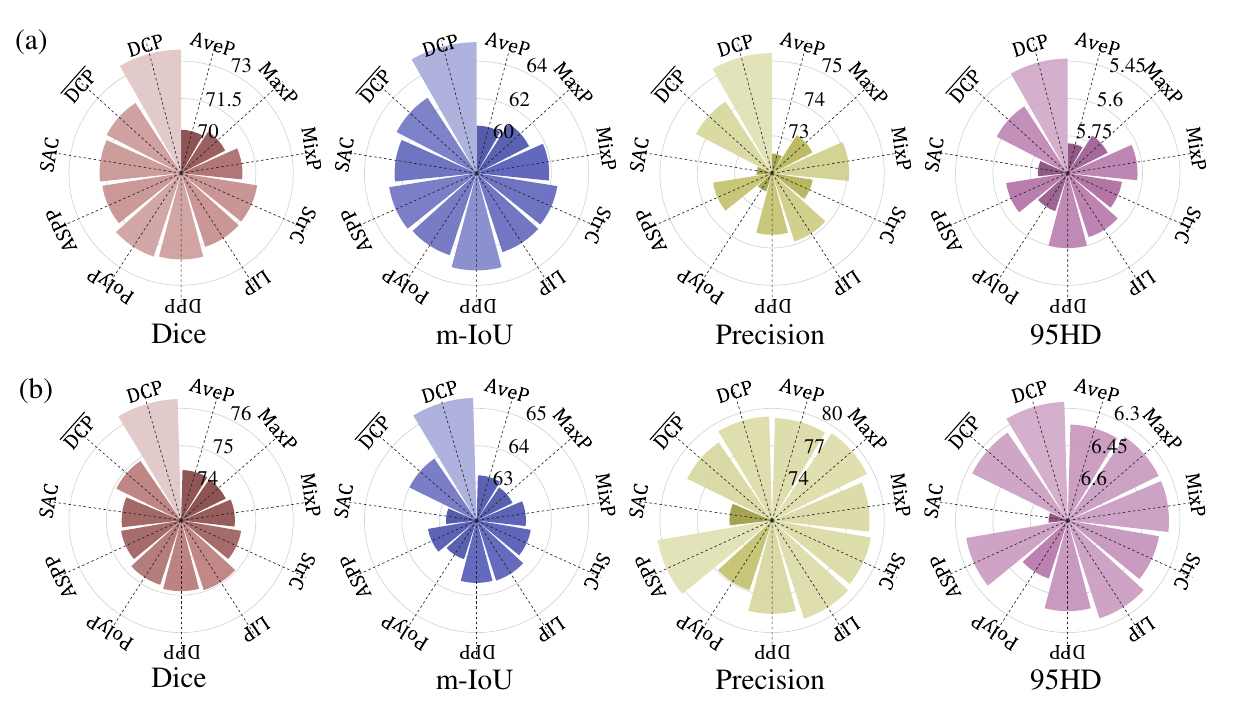}
		\caption{Circular barplots of four evaluation metrics achieved by DC-Net and its variants (for validating the efficacy of DCP operator) on (a) BUS and (b) TUS datasets.}
		\label{fig_6}
		\vspace{-5mm}
	\end{figure}

\begin{figure*}[!t]
		\centering
		\includegraphics[width=1\linewidth]{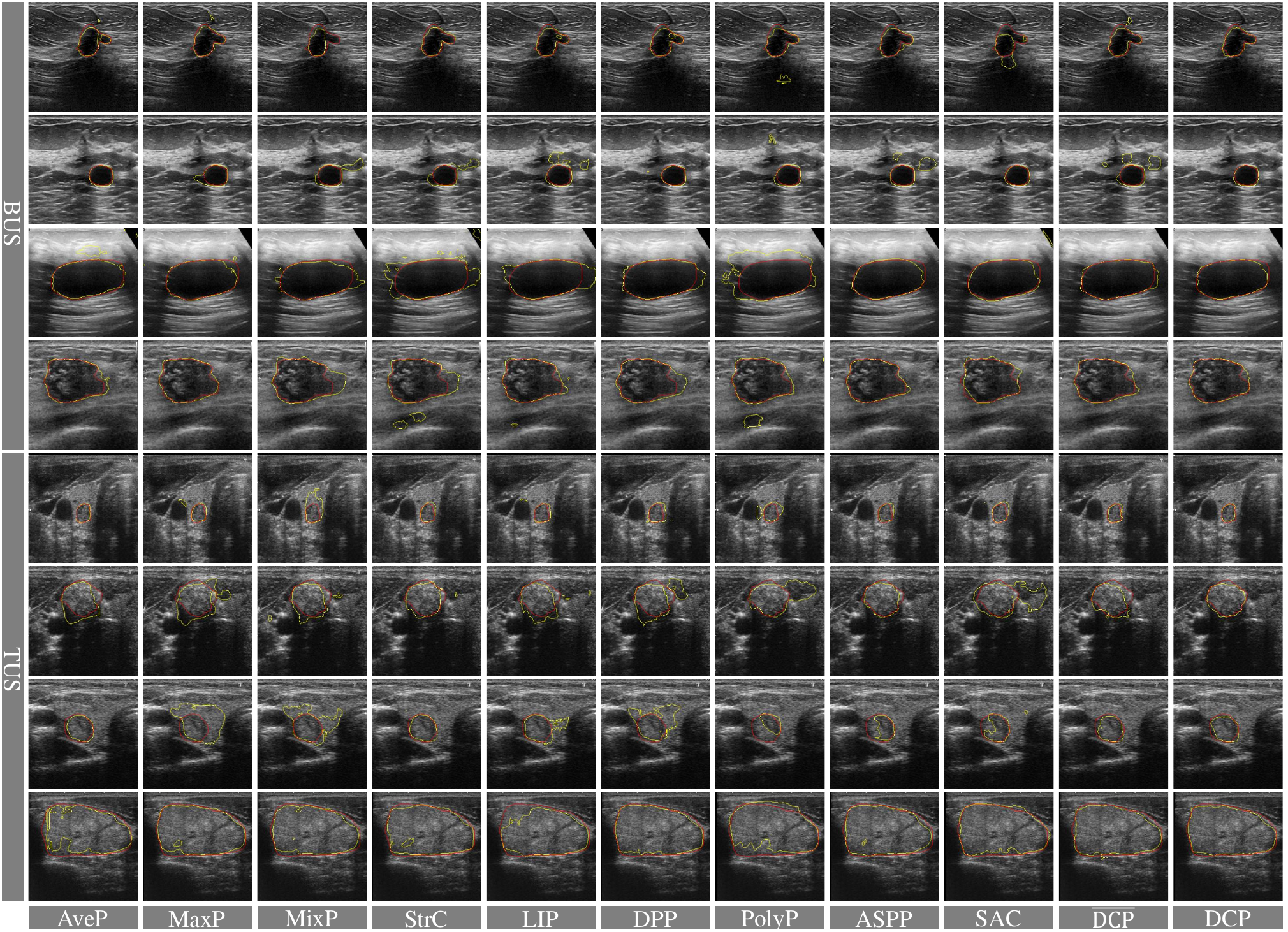}
		\caption{Visualization results of different pooling strategies on some representative cases from BUS and TUS datasets. The red and yellow curves represent the ground truth and the segmentation results of different pooling strategies, respectively.}
		\label{figS4}
		\vspace{-5mm}
	\end{figure*}

	We additionally experimented with equipping the network with the following pooling approaches to help specify the efficacy of DCP operator.
	1) Classical pooling methods in lesion segmentation network: average pooling (AveP), max pooling (MaxP) and mix-pooling (MixP);
	2) Local detail-based pooling approaches: stride convolution (StrC), polynomial pooling (PolyP)~\cite{wei2019building}, local importance-based pooling (LIP) \cite{gao2019lip} and detail-preserving pooling (DPP) \cite{saeedan2018detail};
	3) Context aggregation strategies: atrous spatial pyramid pooling (ASPP)~\cite{chen2018encoder} and switchable ASPP (SAC)~\cite{qiao2021detectors}; 
	4) The variant of DCP (denoted as $\overline{\mbox{DCP}}$) that neglects context aggregation.
	From Fig.~\ref{fig_6}, we can observe some key points.
	\textit{First}, the model with MixP achieves the best performance among three classical pooling methods, perhaps because a compound pooling mode can remedy the pattern dropout caused by single one.
	\textit{Second}, the models with local detail-based pooling approaches (also including DCP and $\overline{\mbox{DCP}}$) work better than those with three classical pooling methods and context aggregation-based methods in general.
	A potential reason is that these methods aggregate local features via the learnable estimation rather than the prior statistics (e.g., local mean or maximum value).
	\textit{Finally}, the DCP method shows more promising results than others, which may owe to three aspects, including 1) learning the adaptive pooling kernel; 2) preserving local contextual details; 3) enlarging the receptive field.
	Fig.~\ref{figS4} gives the visualization results of variants with various pooling strategies on some representative cases from BUS and TUS datasets.
	Obviously, our proposed DCP can reduce the error in separating the lesion region from surrounding tissues that shares similar intensity.
	To analyze the complexity of different pooling strategies, we also calculated the trainable parameters, floating point operations (FLOPs) and the inference time of each image for the model with different pooling operators. All results are reported in Table ~\ref{table_complexity_dcp}. From Table ~\ref{table_complexity_dcp},we can observe that the increasing model parameters with DCP operator is marginal when compared with other operators, which suggests the proposed DCP is relatively efficient.

	\begin{table}[!t]
		\setlength{\abovecaptionskip}{0.cm}
		\renewcommand{\arraystretch}{1.3}
		\caption{Comparison on the complexity of our methods with the variants of different pooling strategies.}
		\label{table_complexity_dcp}
		\centering
		\scriptsize
		\setlength{\tabcolsep}{.3mm}{
			\begin{tabular}{c|c|c|c|c|c|c|c|c|c|c|c}
				\hline	
				Methods	&AveP	&MaxP&	MixP&	StrC&	DPP&	LIP&	ASPP&	SAC	&PolyP&	$\overline{\text{DCP}}$	&DC-Net\\\hline
				Params (M) & 54.4	 & 54.0  & 54.4 & 55.2 & 54.4 & 55.2 & 63.4 & 54.1 & 54.4	& 54.4  & 54.4 \\
				FLOPs (M)	&108.9	&107.9	&108.9	&110.4	&108.9	&110.4	&126.7	&108.3	&108.8	&108.9	& 108.9 \\
				Infer (ms)     	&0.107	&0.107	&0.107	&0.109	& 0.111	&0.116	&0.120	&0.077	&0.071	&0.110	&0.135	\\\hline	
		\end{tabular}}
		\vspace{-5mm}
	\end{table}
	
	\begin{figure}
		\centering
		\includegraphics[width=1\linewidth]{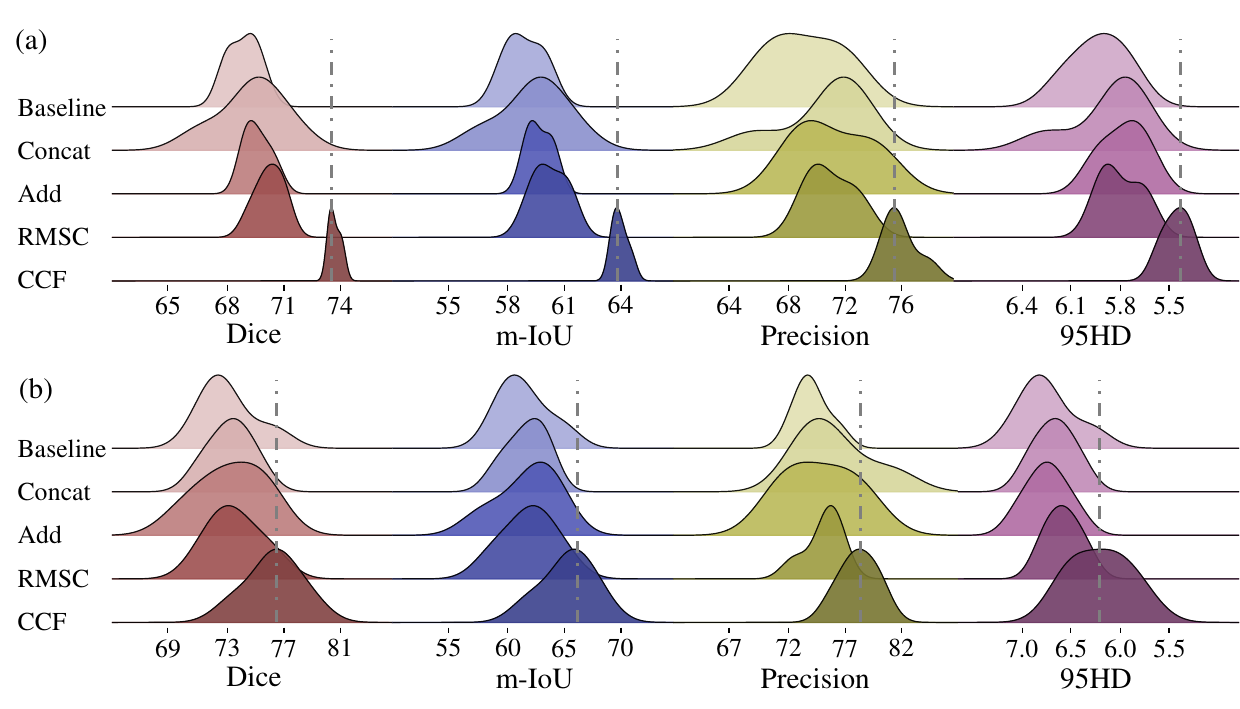}
		\caption{Ridge plots of four evaluation metrics achieved by DC-Net and its variants (for validating the efficacy of CCF module) on (a) BUS and (b) TUS datasets.}
		\label{fig_7}
		\vspace{-5mm}
	\end{figure}
	
	\begin{figure}[h]
		\centering
		\includegraphics[width=1\linewidth]{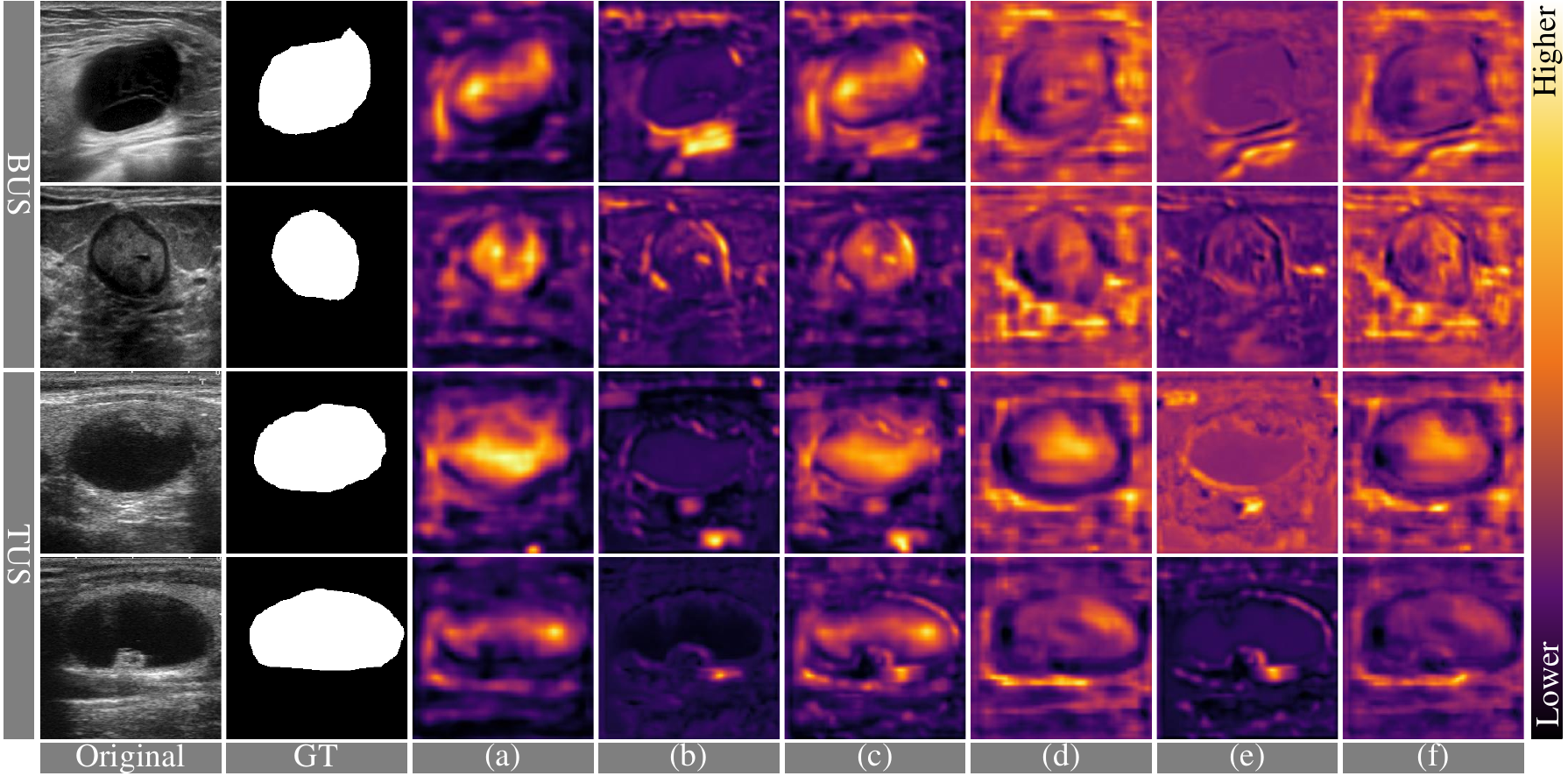}
		\caption{Visualization results of the integrated (via pixel-wise addition operation) low-level and high-level features of CCF module on some representative cases from BUS and TUS datasets. (a)$\thicksim$(f) represent the summed feature maps of $\textbf{X}_E$, $\hat{\textbf{X}}_D$, $\textbf{X}_E\oplus\hat{\textbf{X}}_D$, $\bar{\textbf{X}}_E$, $\bar{\textbf{X}}_D$ and $\bar{\textbf{X}}_E\oplus\bar{\textbf{X}}_D$, respectively.}
		\label{fig_ccf_wa}
		\vspace{-5mm}
	\end{figure}
	
	\subsection{Efficacy of CCF Module}
\begin{figure*}[!t]
		\centering
		\includegraphics[width=1\linewidth]{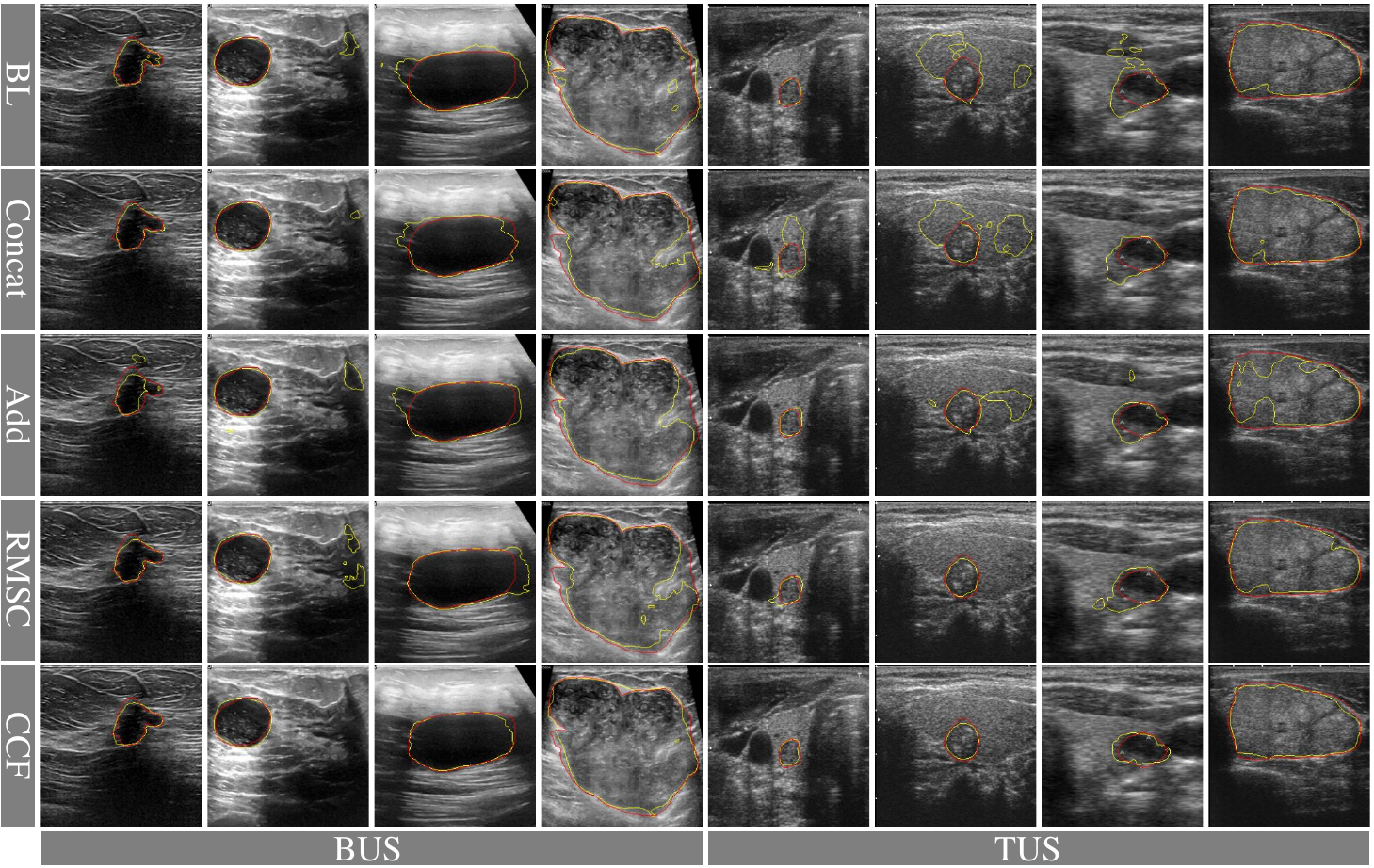}
		\caption{Visualization results of various skip connection strategies on some representative cases from BUS and TUS datasets. The red and yellow curves represent the ground truth and the segmentation results of different skip connection strategies, respectively. Note that "BL" means the Baseline.}
		\label{figS5}
		\vspace{-5mm}
	\end{figure*}

	This ablation study is to validate the effectiveness of CCF module by comparing it with other methods.
	Concretely, these methods include: 
	1) Baseline: it removes the CCF from DC-Net;
	2) Concat/Add/RMSC: it replaces the CCF with the channel-wise concatenation operation, channel-wise addition operation and residual multi-scale connection strategy \cite{honghan2021rms}, respectively.
	From Fig.~\ref{fig_7}, we can observe that the model with Concat or Add performs better than Baseline.
	It demonstrates the skip connection strategy that transmits low-level features from encoder to supplement morphological information for decoder contributes to the performance improvement by a large amount.
	It is worthwhile mentioning that the model with RMSC overpasses those with Concat and Add on two tasks.
	The performance gain may benefit from the multi-scale skip fusion strategy and attention mechanism.
	In addition, CCF is superior to other competing methods, which implies that fusing low-level visual information with high-level semantic representations under the prior of correlation can reduce redundant information and strengthen information connection.
	Fig.~\ref{fig_ccf_wa} displays the integrated (via pixel-wise addition operation) low-level and high-level feature maps (i.e., $\textbf{X}_E$, $\hat{\textbf{X}}_D$, $\textbf{X}_E\oplus\hat{\textbf{X}}_D$, $\bar{\textbf{X}}_E$, $\bar{\textbf{X}}_D$ and $\bar{\textbf{X}}_E\oplus\bar{\textbf{X}}_D$) from CCF module. 
	As shown in Fig.~\ref{fig_ccf_wa}, low-level feature maps contain rich textural information (see Fig.~\ref{fig_ccf_wa}(a)(d)), while high-level ones have better capability of lesion localization (see Fig.~\ref{fig_ccf_wa}(b)(e)).
	Also, the prior of channel correlation can enhance the network's ability of capturing detail information, such as shape and boundary (as shown in Fig.~\ref{fig_ccf_wa}(a)(b) and (d)(e)).
	Therefore, feature integration under the prior of channel correlation can strengthen the efficiency of feature propagation.
	Fig.~\ref{figS5} displays the visualization results to qualitatively compare various skip connection strategies.
	Though all methods can accurately locate the lesion, the model with CCF shows better ability in fitting the lesion shape, regardless of it with large-size or small-size. 
	
	\begin{figure}[h]
		\centering
		\includegraphics[width=1\linewidth]{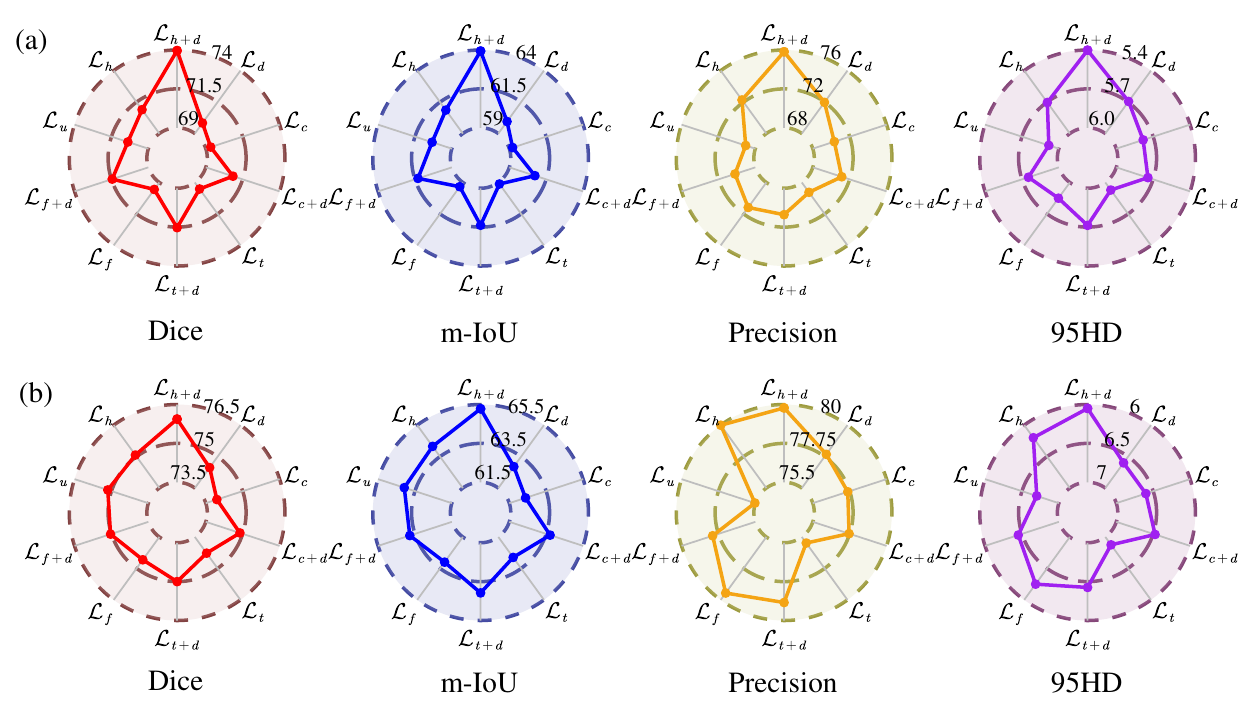}
		\caption{Radar plots of four evaluation metrics achieved by DC-Net and its variants (for validating the efficacy of harmonic loss) on (a) BUS and (b) TUS datasets.}
		\label{fig_8}
		\vspace{-5mm}
	\end{figure}
	
	\begin{figure*}[!t]
		\centering
		\includegraphics[width=1\linewidth]{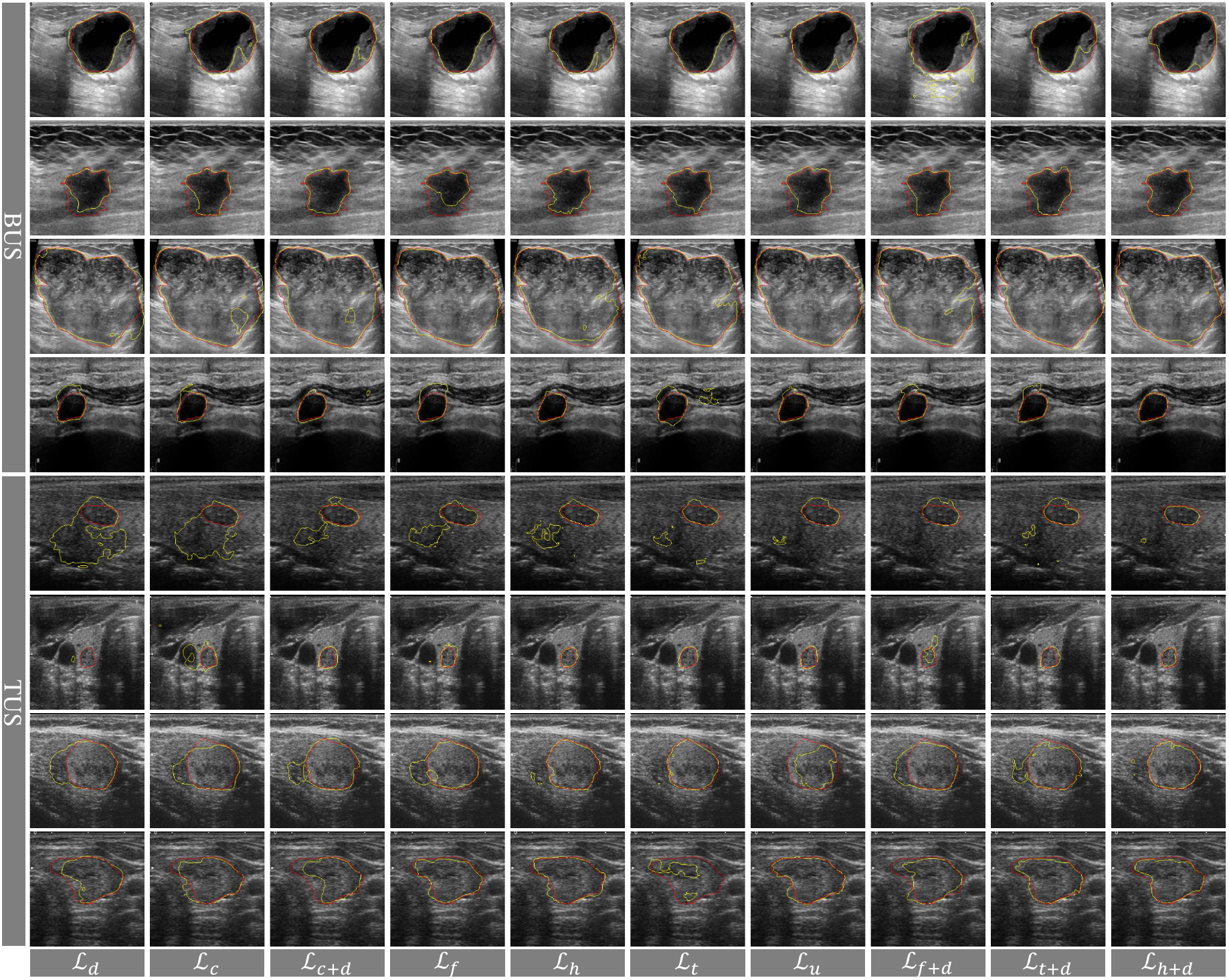}
		\caption{Visualization results of all optimization functions on some representative cases from BUS and TUS datasets. The red and yellow curves represent the ground truth and the segmentation results of different optimization functions, respectively.}
		\label{figS6}
		\vspace{-5mm}
	\end{figure*}
	
	\subsection{Efficacy of Optimization Function}
	Here, we mainly make a comparison the proposed optimization loss (i.g., harmonic loss $\mathcal{L}_h$ and its hybrid version with Dice loss $\mathcal{L}_{h+d}$) with the following losses:
	1) Single function: CE loss $\mathcal{L}_c$, Dice loss $\mathcal{L}_d$, focal loss $\mathcal{L}_f$, topk loss $\mathcal{L}_{t}$~\cite{wu2016bridging};
	2) Hybrid function: joint CE loss and Dice loss $\mathcal{L}_{c+d}$, generalized Dice and CE-based loss $\mathcal{L}_u$ \cite{yeung2022unified}, joint focal loss and Dice loss $\mathcal{L}_{f+d}$, joint topk loss and Dice loss $\mathcal{L}_{t+d}$.
	Fig.~\ref{fig_8} shows the experimental results.
	Generally speaking, the models trained with hybrid loss functions work better compared to those trained with single loss functions, which further verifies the conclusion stated by \cite{MA2021102035lossodyssey}.
	As expected, $\mathcal{L}_c$ is inferior to its generalized variants (i.e., $\mathcal{L}_f$, $\mathcal{L}_u$, $\mathcal{L}_h$, and $\mathcal{L}_t$), as the latter compels the network to give more attention to hard samples during the training stage.
	\begin{figure*}
		\centering
		\includegraphics[width=1\linewidth]{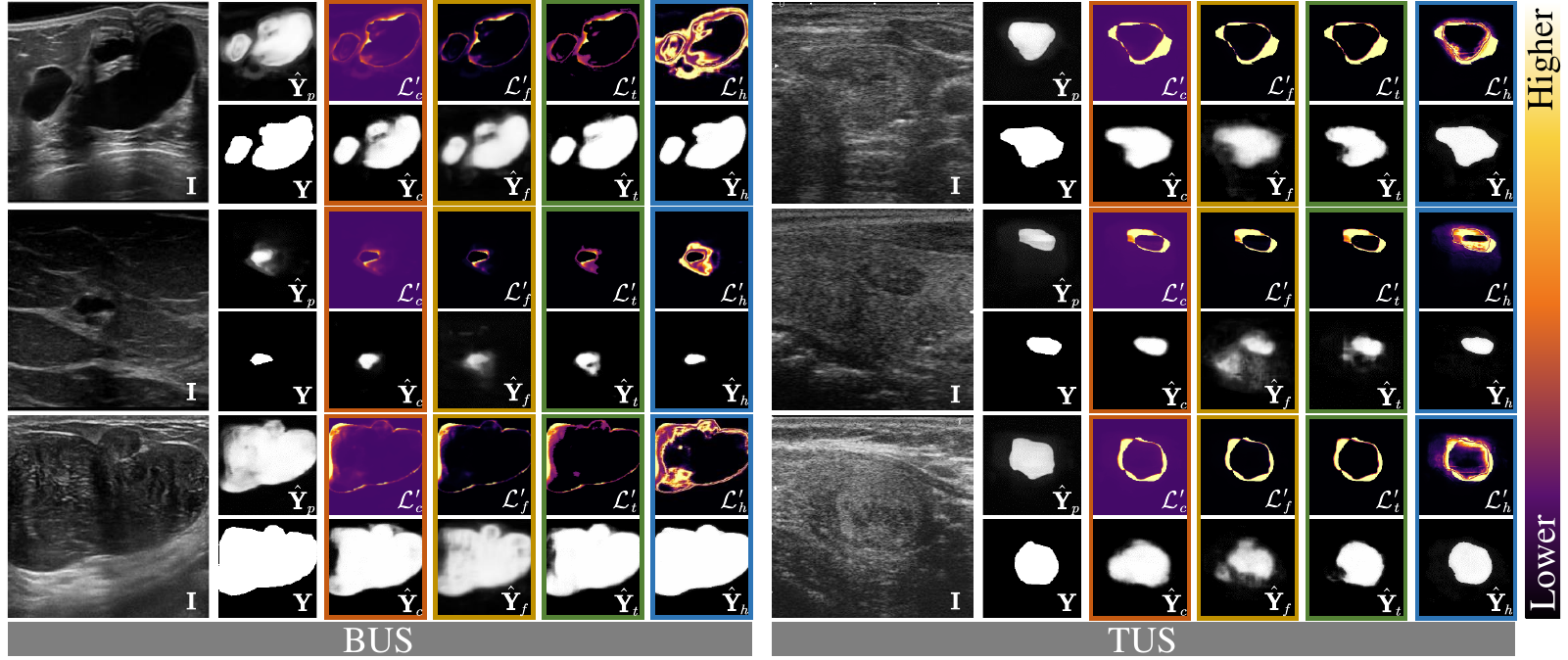}
		\caption{Gradient maps of several representative losses (i.e., $\mathcal{L}'_{c}$, $\mathcal{L}'_{f}$, $\mathcal{L}'_{t}$, and $\mathcal{L}'_{h}$) on the same prediction map (i.e., $\hat{\textbf{Y}}_p$) from previous training epoch. For convenience, $\hat{\textbf{Y}} _{c}$, $\hat{\textbf{Y}}_{f}$, $\hat{\textbf{Y}}_{t}$, and $\hat{\textbf{Y}}_{h}$ denote the final converged prediction results that are matched with the corresponding loss functions. $\textbf{I}$ and $\textbf{Y}$ represent the original image and ground truth, respectively.}
		\label{fig_loss_grad}
		\vspace{-7mm}
	\end{figure*}
	The models with $\mathcal{L}_h$ and $\mathcal{L}_{h+d}$ achieve the highest metric scores among those models with single functions and hybrid functions, respectively.
	Unlike $\mathcal{L}_f$, $\mathcal{L}_u$ and $\mathcal{L}_{t}$, the proposed $\mathcal{L}_h$ can down-weight easy samples, preserve and even lift the loss margin and gradient of hard samples.
	More importantly, it can also narrow down the suppression range to prevent low-confidence samples from being overwhelmed.
	Fig.~\ref{figS6} visualizes some representative cases from BUS and TUS datasets for qualitative comparison of various loss functions.
	It can be found that the output of model trained with $\mathcal{L}_{h+d}$ can approximately coincide with the ground truth.
	Meanwhile, we also notice $\mathcal{L}_{h+d}$ helps model work in the blurry-boundary scenario, especially on TUS datasets.
	Fig.~\ref{fig_loss_grad} shows the gradient maps of several representative losses on the same prediction map from previous training epoch.
	We mainly focus on the comparison of cross-entropy loss $\mathcal{L}_c$ and its advanced variants (i.e., focal loss $\mathcal{L}_f$, topk loss $\mathcal{L}_t$ and harmonic loss $\mathcal{L}_h$) to analyze their influence on easy, low-confidence and hard samples.
	From Fig.~\ref{fig_loss_grad}, we have the following findings.
	1) $\mathcal{L}_c$ still produces gradients for easy samples/pixels (e.g., purple pixels in $\mathcal{L}'_c$).
	Although these gradients are weak, they may dominant and affect the direction of network optimization due to the large proportion of easy samples in the whole image.
	2) Compared with $\mathcal{L}_c$, $\mathcal{L}_f$ can obviously suppress the gradient of easy samples (e.g., those pixels changes from purple to black), but it also suppresses the low-confidence ones (e.g., $\mathcal{L}'_f$ has more incomplete contours of some lesions in BUS images than $\mathcal{L}'_c$).
	3) $\mathcal{L}_t$ relatively alleviates the aforementioned issues, but it still pays insufficient attention to low-confidence samples, which results in flawy boundaries of some lesions.
	4) $\mathcal{L}_h$ can not only suppress the gradient of easy samples, but also focus on the hard and low-confidence samples around the boundary (e.g., yellow and purple pixels in $\mathcal{L}'_h$), so as to obtain promising segmentation performance.

	\section{Discussion}
	\label{Discussion}
	\subsection{Comparison with Saliency Map-based Methods}
	In this subsection, we compare the proposed DC-Net with several saliency map-based methods, including iFCN~\cite{xu2016deep}, DeepIGeoS~\cite{wang2018deepigeos}, and SMU-Net~\cite{ning2021smu}.
	All of them require interactive saliency map generation. 
	For deeper comparison, we run the DC-Net by using the interactively-generated saliency maps from other methods as input, and also feed the automatically-generated saliency map from DC-Net into other methods.
	For convenience, we mark those methods based on interactively-generated and automatically-generated saliency maps with the tails of "interac" and "automac", respectively.
	Experimental results are shown in Table~\ref{table_semi-auto}.
	From Table~\ref{table_semi-auto}, we have several observations as follows.
	1) DC-Net demonstrates enhanced performance when utilizing interactively-generated saliency maps from SMU-Net. 
	But, it experiences significant performance degradation when employing interactively-generated saliency maps from DeepIGeoS.
	The saliency maps from SMU-Net cover low-level and high-level image structures that help distinguish lesion from background.
	However, subtle gray difference between lesion and background leads to the blurry boundary in the geodesic distance map (i.e., interactively-generated saliency maps from DeepIGeoS), which may give wrong prompt and hinders the performance improvement.
	Even so, both of them require interactive hand-clicked seeds, thus are time-consuming and experience-dependent.
	It is worth mentioning that the proposed method has smaller model parameters than SMU-Net (54.4M \textit{vs.} 103.3M).
	2) When the same saliency maps are used, DC-Net can achieve comparable and even higher results than others, which demonstrates its efficacy.

	\begin{table}[!t]
		\setlength{\abovecaptionskip}{0.cm}
		\renewcommand{\arraystretch}{1.1}
		\caption{Segmentation results (mean $\pm$ standard deviation) of DC-Net and saliency map-based methods on BUS and TUS datasets.}
		\label{table_semi-auto}
		\centering
		\scriptsize
		\setlength{\tabcolsep}{1mm}{
			\begin{tabular}{c|c|cccc}
				\hline	
				& Methods           & Dice                & m-IoU             & Precision         & 95HD			\\ \cmidrule{1-6}
				
				\multirow{12}{*}{BUS}   
				& DeepIGeoS (interac)	& 64.46$\pm$1.47      & 53.95$\pm$1.70	  & 61.59$\pm$1.93    & 6.57$\pm$0.18 \\
				& DeepIGeoS (automac)   & 65.07$\pm$2.04      & 54.60$\pm$2.28    & 62.88$\pm$2.67    & 6.45$\pm$0.30  \\
				& DC-Net (interac)     & 67.60$\pm$2.23      & 57.98$\pm$1.92    & 68.14$\pm$0.35    & 5.94$\pm$0.05 \\
				& DC-Net (automac)      & 73.96$\pm$0.36	  & 64.26$\pm$0.48	  & 73.95$\pm$0.36	  & 5.44$\pm$0.10 \\\cmidrule{2-6}
				& iFCN (interac)	    & 73.98$\pm$0.59      & 63.12$\pm$0.80     & 71.44$\pm$1.22    & 5.71$\pm$0.09 \\
				& iFCN (automac)        & 69.50$\pm$0.79      & 59.15$\pm$0.82    & 69.59$\pm$1.08    & 5.83$\pm$0.08\\
				& DC-Net (interac)     & 74.18$\pm$0.93      & 64.01$\pm$1.08    & 75.75$\pm$2.29    & 5.54$\pm$0.15 \\
				& DC-Net (automac)      & 73.96$\pm$0.36	  & 64.26$\pm$0.48	  & 73.95$\pm$0.36	  & 5.44$\pm$0.10 \\\cmidrule{2-6}
				& SMU-Net (interac)    & 80.81$\pm$0.87      & 71.20$\pm$1.05     & 80.49$\pm$1.89    & 4.96$\pm$0.08 \\
				& SMU-Net (automac)     & 73.87$\pm$0.65      & 63.40$\pm$0.84    & 74.37$\pm$1.26    & 5.52$\pm$0.02 \\
				& DC-Net (interac)     & 80.76$\pm$0.17      & 71.17$\pm$0.30     & 83.98$\pm$0.95    & 4.93$\pm$0.02 \\
				& DC-Net (automac)      & 73.96$\pm$0.36	  & 64.26$\pm$0.48	  & 73.95$\pm$0.36	  & 5.44$\pm$0.10 \\\cmidrule{1-6}
				\multirow{12}{*}{TUS}   & DeepIGeoS (interac)  & 71.97$\pm$0.47      & 59.62$\pm$0.98    & 75.12$\pm$2.44    & 6.68$\pm$0.22 \\
				& DeepIGeoS (automac)   & 72.98$\pm$1.05      & 60.81$\pm$1.53    & 75.03$\pm$2.84    & 6.63$\pm$0.18 \\
				& DC-Net (interac)     & 70.19$\pm$2.33      & 57.61$\pm$2.38    & 74.33$\pm$1.74    & 6.78$\pm$0.12 \\
				& DC-Net (automac)	    & 76.25$\pm$1.89	  & 65.28$\pm$2.24	  & 76.25$\pm$1.89	  & 6.22$\pm$0.26 \\\cmidrule{2-6}
				& iFCN (interac)	    & 74.02$\pm$1.98	  & 60.99$\pm$2.53	  & 76.24$\pm$2.85	  & 6.64$\pm$0.22 \\
				& iFCN (automac)        & 67.15$\pm$1.40      & 54.40$\pm$1.43    & 67.22$\pm$2.86    & 7.05$\pm$0.18 \\
				& DC-Net (interac)     & 74.72$\pm$1.90      & 62.02$\pm$2.41    & 77.19$\pm$3.42    & 6.64$\pm$0.11 \\
				& DC-Net (automac)	    & 76.25$\pm$1.89	  & 65.28$\pm$2.24	  & 76.25$\pm$1.89	  & 6.22$\pm$0.26 \\\cmidrule{2-6}
				& SMU-Net (interac)	& 76.34$\pm$1.77      & 63.91$\pm$2.16	  & 77.37$\pm$2.72	  & 6.40$\pm$0.20 \\
				& SMU-Net (automac)     & 72.84$\pm$2.47      & 59.59$\pm$2.84    & 75.13$\pm$2.85    & 6.62$\pm$0.19 \\
				& DC-Net (interac)     & 80.76$\pm$1.87      & 69.74$\pm$2.41    & 83.38$\pm$2.25    & 6.40$\pm$0.20 \\
				& DC-Net (automac)	    & 76.25$\pm$1.89	  & 65.28$\pm$2.24	  & 76.25$\pm$1.89	  & 6.22$\pm$0.26 \\\hline
		\end{tabular}}
		\vspace{-5mm}
	\end{table}
	
	\subsection{Parameters Analysis}
\begin{figure}
		\centering
		\includegraphics[width=1\linewidth]{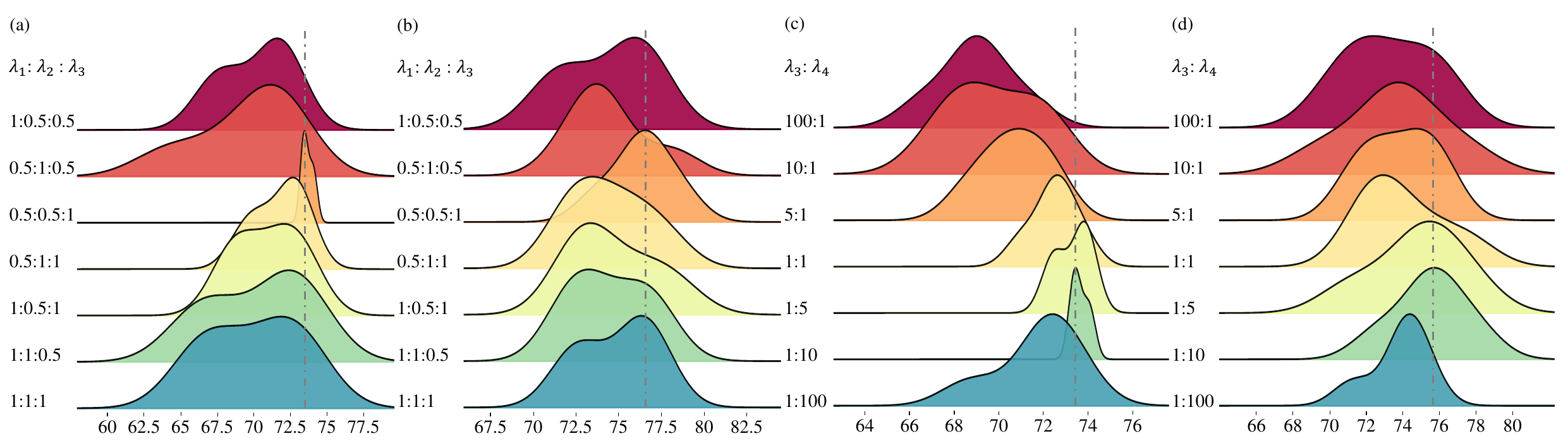}
		\caption{Sierra plots of Dice score achieved by DC-Net with different parameters ($\lambda_1$, $\lambda_2$, $\lambda_3$ in Eq.(18) and $\lambda_4$ in Eq.(27)) on (a)/(c) BUS and (b)/(d) TUS datasets.}
		\label{FigS8}
		\vspace{-2mm}
	\end{figure}

\begin{figure}[!t]
		\centering
		\includegraphics[width=1\linewidth]{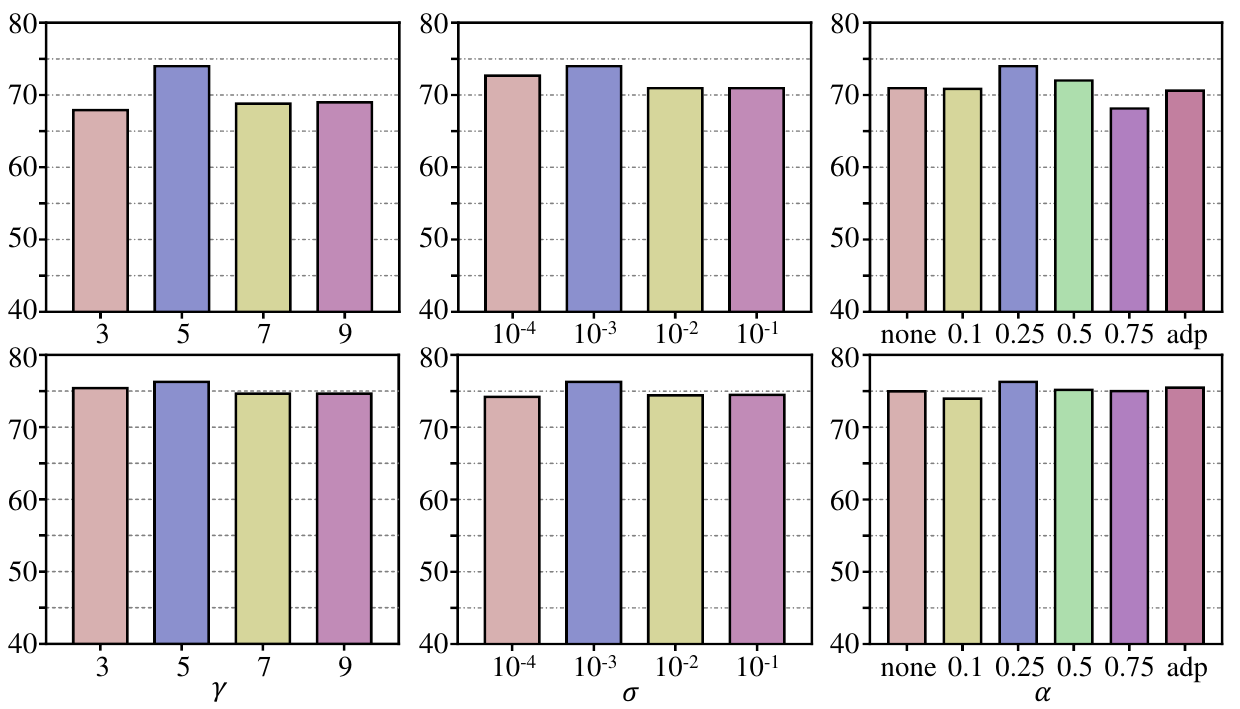}
		\caption{Bar plots of Dice score achieved by DC-Net with different parameters ($\gamma$, $\sigma$ and $\alpha$ in Eq.(26)). From top to bottom are BUS and TUS datasets, repsectively.}
		\label{FigS9}
		\vspace{-4mm}
	\end{figure}

	In this part, we discuss the influence of key parameters in our proposed method, including $\lambda_1$, $\lambda_2$, $\lambda_3$ in Eq.(18), $\lambda_4$ in Eq.(27), and $\gamma$, $\sigma$, $\alpha$ in Eq.(26).
	1) To determine the ratio of $\lambda_1$, $\lambda_2$, $\lambda_3$, we tested the model under different settings on two tasks.
	As shown in Fig.~\ref{FigS8} (a)-(b), the model gets the best performance when the ratio is set to 0.5: 0.5: 1.
	It is not surprising about that as focusing more attention on final precise segmentation (via the couple subnet) is beneficial to performance improvement.
	We further investigate the ratio of $\lambda_3$ and $\lambda_4$.
	From Fig.~\ref{FigS8} (c)-(d), as we expected, a large $\lambda_4$ is a better choice, which assigns a large weight to the harmonic loss.
	And the best results are obtained when the ratio equals to 1:10.
	2) We conducted experiments by varying the value of $\gamma$, $\sigma$ and $\alpha$ in the range of $\{3,5,7,9\}$, $\{10^{-4}, 10^{-3}, 10^{-2}, 10^{-1}\}$ and $\{0.1, 0.25,0.5, 0.75\}$, respectively, for having an insight into their influence on the proposed model.
	In addition, we also designed two unique settings for $\alpha$, including 1) removing $\alpha$ from the harmonic loss (denoted as "none") and 2) adaptively determining $\alpha$ according to the proportion of the foreground to the whole image (denoted as "adp").
	From Fig.~\ref{FigS9}, we can observe that the model with $\gamma=5$, $\sigma=10^{-3}$ and $\alpha=0.25$ outperforms those with other settings, for both tasks.
	As illustrated in Fig.~\ref{FigS9}, the harmonic loss with $\gamma=5$ and $\sigma=10^{-3}$ can suppress easy samples (with $p_t$ in the range of (0.75, 1]) and enhance the gradient of low-confident samples (with $p_t$ in the range of [0.4, 0.75]).
	
	\subsection{Complexity Analysis}
	To discuss the model's complexity, we calculated the number of model parameters, FLOPs and the inference time of each image for several representative algorithms, including DeepLabV3+, U-Net, U-Net++, AU-Net, DeepIGeoS, iFCN, and SMU-Net.
	Table~\ref{table_complexity} shows all comparison results. 
	The complexity of DC-Net approximates that of DeepLab V3+, but is higher than most of other methods.
	It is worth mentioning that both SMU-Net and DC-Net contain special streams or branches to deal with foreground and background representation, and they experience a significant performance improvement at the cost of a relatively high complexity.
	But, compared to SMU-Net, DC-Net seems to be more efficient and is free of manual interaction.

\section{Conclusion}
	In this paper, we present a decomposition-coupling network, referred to as DC-Net, for lesion segmentation in complex-scenario ultrasound images by integrating saliency map generation and lesion segmentation into a collaborative framework.
	We devise a differentiable context pooling operator and a cross-correlation fusion module for dimension reduction and resolution restoration, respectively.
	And a novel coupler is proposed to conduct dependency-aware reinforcement of foreground-salient representation with the complementary background information.
	We further introduce a harmonic loss function to take into account the efficient optimization of low-confidence samples in additional to hard ones.
	Extensive experiments and evaluations on two ultrasound lesion segmentation tasks have demonstrated the proposed method outperforms several recent deep learning methods.
	
	\begin{table}[!t]
		\setlength{\abovecaptionskip}{0.mm}
		\renewcommand{\arraystretch}{1.3}
		\caption{The model's complexity of DC-Net and several representative algorithms.}
		\label{table_complexity}
		\centering
		\scriptsize
		\setlength{\tabcolsep}{0.3mm}{
			\begin{tabular}{c|c|c|c|c|c|c|c|c}
				\hline
				Methods & DeepLabV3+ & U-Net & U-Net++ & AU-Net & DeepIGeoS & iFCN & SMU-Net & DC-Net  \\ \hline  
				Params (M) & 41.0 & 31.4 & 23.8 & 34.9 & 1.64 & 14.7 & 103.3  & 54.4    \\
				FLOPs (M)	 & 163.7	 & 62.8	 & 47.5	&69.8	 & 3.28 & 29.5	& 206.7	& 108.9 \\
				Infer (ms) & 0.052 & 0.097 & 0.102 & 0.104	&0.251	&0.126 & 0.124 & 0.135 \\\hline	
		\end{tabular}}
		\vspace{-5mm}
	\end{table}

	\bibliography{IEEEabrv,reference}
	\bibliographystyle{IEEEtran}
	
\end{document}